\documentclass[]{jfm}
\usepackage{graphicx}
\usepackage{newtxtext}
\usepackage{newtxmath}
\usepackage{natbib}
\usepackage{hyperref}
\hypersetup{
    colorlinks = true,
    urlcolor   = blue,
    citecolor  = black,
}

\newcommand{\RomanNumeralCaps}[1]
\linenumbers

\usepackage{subfigure}

\title{Linear and nonlinear optimal growth mechanisms for generating turbulent bands}

\author{E.Parente\aff{1,2} \corresp{\email{enzaparente@gmail.com}},
\and J.-Ch. Robinet \aff{1}, P. De Palma\aff{2}
\and S. Cherubini\aff{2}}

\affiliation{\aff{1} DynFluid - Arts et M\'{e}tiers Paris, 151 Bd de l'H\^{o}pital, 75013 Paris, France
\aff{2} Dipartimento di Meccanica, Matematica e Management (DMMM), Politecnico di Bari, Via Re David 200, 70126 Bari, Italy}

\begin{document}
\maketitle

\begin{abstract}
Linear and nonlinear energy optimizations in a tilted domain are used to unveil the main mechanisms allowing the creation of a turbulent band in a channel flow. Linear optimization predicts an optimal growth for streamwse and spanwise wavenumbers $k_x = 1.2$, $k_z = - 1.75$, corresponding to the peak values of the premultiplied energy spectra of direct numerical simulations. At target time, the linear optimal perturbation is composed by oblique streaks, which, for a sufficiently large initial energy, induce turbulence in the whole domain, due to the lack of spatial localization. When localization is achieved by adding nonlinear effects to the optimization, or by artificially confining the linear optimal to a localized region in the spanwise direction, a large-scale flow is created, which leads to the generation of a localised turbulent band. These results suggest that inducing transition towards turbulent bands in a tilted domain, two main elements are needed: a linear energy growth mechanism such as the lift-up for generating large-amplitude flow structures which produce inflection points; large-scale vortices ensuring spatial localisation. Remarkably, these two elements alone are able to generate an isolated turbulent band also in a large, non-tilted domain.
\end{abstract}

%\begin{keywords}
%Authors should not enter keywords on the manuscript, as these must be chosen by the author during the online submission process and will then be added during the typesetting process (see \href{https://www.cambridge.org/core/journals/journal-of-fluid-mechanics/information/list-of-keywords}{Keyword PDF} for the full list).  Other classifications will be added at the same time.
%\end{keywords}

%{\bf MSC Codes }  {\it(Optional)} Please enter your MSC Codes here

\section{Introduction}
In plane Poiseuille flow, transition to turbulence often arises for Reynolds numbers not surpassing the critical value for linear stability analysis, $Re = 5772$.  For  Reynolds numbers consistently lower than this threshold value, the flow may exhibit localised turbulence \citep{tsukahara2005}.
\cite{carlson1982} was the first to experimentally observe laminar-turbulent patterns in the channel flow, for $Re = 1000$.
In very large domains, localised turbulent bands tilted with respect to the streamwise direction, plunged in the laminar flow, are observed \citep{tsukahara2014,xiong2015,tao2018,shimizu2019,kashyap2020a}. This oblique laminar-turbulent pattern is also observed in other shear flows, although for a different range of Reynolds numbers and  presenting different angles \citep{prigent2002,barkley2005,duguet2010,tuckerman2011,duguetPRL2013,chantry2017,tuckerman2020}.  Using numerical simulations, \cite{tao2018} have established that oblique turbulent bands may arise in the channel flow at $Re \approx 660$. Recently, forcing the flow with an inflectional instability, \cite{song2020} have been able to generate turbulent bands at $Re\approx500$, although not self-sustained.
Turbulent bands have been analysed by several authors in large domains with the aim of studying their characteristics, such as their angle and length \citep{tsukahara2005,tao2018,kashyap2020a}, as well as their dynamics and interactions \citep{duguet2010,tao2018,shimizu2019,gome2020}. In order to reduce the computational cost and the flow complexity, \cite{barkley2005,barkley2007} studied the behaviour of the plane Couette flow in a small domain tilted perpendicularly to the turbulent band direction. Later,  \cite{tuckerman2014} extended this methodology to  the plane Poiseuille flow. \\
Recently, many works have focused on the origin and growth of turbulent bands. In a large domain, \cite{shimizu2019} observed that streaks are generated at the head of the turbulent band and suggest that streak generation could be the origin of the self-sustaining process of a single turbulent band. According to this hypothesis, investigating a small domain at the head of the band, \cite{xiao2020} have performed a linear stability analysis %and a linear optimization 
of the mean flow computed in three different  regions at the head of the band. They have notably found that an inflectional spanwise instability generates streaky structures similar to those found at the head of a turbulent band, and they have proposed that this instability can be at the origin of the growth of the turbulent band. Based on this hypothesis, \cite{song2020} have searched for a forcing that induces inflectional instability in the flow able to trigger turbulence in the form of turbulent bands. They found a fast non-modal growth associated with the base velocity profile deformed by this continuous forcing. On the other hand, using a nonlinear approach, \cite{paranjape2020} have searched for an  edge state in a tilted domain and they have found a localised nonlinear travelling wave solution that shows  properties very similar to those of turbulent bands in a tilted domain.
Conversely, \cite{tao2018} suggest that, for triggering and sustaining a turbulent band, a large-scale flow is necessary.
%, due to the two flow scales that characterised the turbulent-laminar pattern: a small scale flow and a large scale flow. 
In fact, all turbulent bands are characterized by a small-scale flow inside the turbulent region,  characterised by streaks and vortices, surrounded by large-scale vortices,  parallel to the turbulent bands and having opposite direction on their two sides. In particular, \cite{duguetPRL2013} have argued that the validity of the continuity equation for this large-scale flow is responsible for the turbulent band oblique evolution.
\\
\cite{shimizu2019} %kanazawa2018}
observed that turbulent bands are sustained by an active streak generation at the head of the bands, while streaks decay is found in the tail.  Moreover, \cite{tao2018} reported a strong increase of the total disturbance kinetic energy corresponding with the creation of turbulent bands. This energy increase follows an almost algebraic growth, instead of an exponential behaviour, as should be expected in the case of asymptotic instability as that reported by \cite{xiao2020}. It is known since the pioneer work of \cite{landahl1980note}, that an algebraic kinetic energy growth of perturbations is induced when weak counter-rotating vortices generate high-amplitude streaky structures by means of the lift-up mechanism. These two flow structures, which are easily retrieved by optimal transient growth analysis  \citep{Luchini2000},  are two of the fundamental elements of the self-sustaining process which supports turbulence in shear flows \citep{Hamilton1995,Waleffe1997}, together with secondary instability of the streaks which is linked to the creation of inflectional points in the velocity profiles.  
Despite the observation of a consistent kinetic energy growth together with streaks generation in the head of a turbulent band,  the possible relation between  turbulent bands generation, the optimal energy growth of streaks and their inflectional instability, have still not been investigated in the literature. 
\\
The aim of this study is to elucidate the possible link between  transient energy growth mechanisms and the generation and sustaining of turbulent bands in channel flow. For investigating the energy growth mechanism in detail, we restrain the  analysis to a tilted domain allowing the generation of a single localised turbulent band, as previously done in direct numerical simulations \citep{tuckerman2014}.
In this study, we search for linear and nonlinear optimal perturbations for the channel flow in a tilted domain at $Re=1000$. We have found that transient growth of streaks is able to generate turbulent bands, although only in the presence of a large-scale flow arising from spatial localization. %In addition, we distinguish the mechanisms linked to the two flow scales. % Thus, the aim of this work is to find the characteristics of an optimal perturbation able to trigger turbulent bands in a tilted domain.
\\
The paper is organised in the following way: the problem formulation is presented in section \ref{sec:problem}. Then, in section \ref{sec:results} the linear and nonlinear optimal perturbations are shown and discussed. Finally, conclusions are drawn in section \ref{sec:conclusion}.

\section{Problem formulation}\label{sec:problem}

For reducing the problem complexity and the computational cost,  a tilted domain is considered  for  analysing oblique turbulent bands in plane Poiseuille flow, as previously done by \cite{barkley2005,barkley2007} for plane Couette flow and by \cite{tuckerman2014} for plane Poiseuille flow. Starting from the classical plane Poiseuille flow, $\textbf{U}_P = [U_P(y), 0, 0]^T$, with $U_P (y) = 1-y^2$, defined in the coordinate system $\mathbf{x}'=(x',y',z')^T$, where $x'$ indicates the direction of the flow $U_P$, the tilted domain is obtained by applying the following change of reference:
\begin{equation*}
    \mathbf{\hat{e}}_{x} = cos \theta \mathbf{\hat{e}}_{x'} - sin \theta \mathbf{\hat{e}}_{z'}, \qquad
    \mathbf{\hat{e}}_{y} = \mathbf{\hat{e}}_{y'}, \qquad
    \mathbf{\hat{e}}_{z} = - sin \theta \mathbf{\hat{e}}_{x'} + cos \theta \mathbf{\hat{e}}_{z'},
\end{equation*}
 $\mathbf{x}=(x,y,z)^T$ being the tilted domain coordinate system, and
$\theta$ being the angle of the new coordinates system, corresponding to the angle of a turbulent band free to evolve in the non-tilted domain.\\
The dynamics of the turbulent bands in the tilted domain can be described by decomposing the instantaneous field into a perturbation $\textbf{u}' = [u',v',w']^T$ and a laminar base flow $\textbf{U} = [U(y), 0, W(y)]^T$, with $U(y) = U_P(y) cos \theta$ and $W(y) = U_P(y) sin \theta$. The perturbation dynamics is governed by the Navier-Stokes equations for incompressible flows, written in a perturbative form with respect to the base flow:
\begin{equation}
\displaystyle \frac{\partial u_i'}{\partial x_i} = 0, \qquad
\displaystyle \frac{\partial u_i'}{\partial t} + u_j'\frac{\partial u_i'}{\partial x_j} + u_j' \frac{\partial U_i}{\partial x_j} + U_j \frac{\partial u_i'}{\partial x_j} = - \frac{\partial p'}{\partial x_i} + \frac{1}{Re} \frac{\partial^2 u_i'}{\partial x_j},
\label{NS}
\end{equation}
with $p'$ the pressure perturbation and $Re = U_{c} h / \nu$ the Reynolds number defined using the centreline velocity of the laminar Poiseuille flow, $U_c$, the half width of the channel, $h$, and the kinematic viscosity $\nu$.\\
\noindent In order to find the optimal solution able to trigger turbulent bands in the tilted domain, we have computed linear and nonlinear optimal perturbations \citep{cherubini2010,pringle2012}. In both cases, we choose as objective function the energy gain $G(T) = E(T)/E(0)$, where $E(t)=1/(2V)\int u_i^2(t) dV$,  $E(T)$ and $E(0)$ being the kinetic energy at the chosen target time and at the initial time, respectively. Thus, we search for the initial perturbation $\textbf{u}'(0)$ providing the largest possible energy  at fixed target time with an optimization loop based on the Lagrange multiplier technique \citep{cherubini2011}.
Linear optimization is carried out using an in-house Matlab code,  whereas nonlinear optimization is implemented within the open source code $Channelflow$ (channelflow.ch) (\cite{channelflow}).\\
For all the simulations the volume flux is kept constant imposing the bulk velocity equal to $U_{bulk} = 2/3$. The same domain size and spatial discretization used by \cite{tuckerman2014} is adopted,  namely $L_x \times L_y \times L_z = 10 \times 2 \times 40$ discretized on $N_x \times N_y \times N_z = 128 \times 65 \times 512$ grid. All computations are performed at $Re = 1000$, for which \cite{tuckerman2014} have reported a persisting turbulent-laminar patter in the form of a single band. % that does not relaminarize. 
The angle of the tilted domain, $\theta$, is chosen equal to $35^{\circ}$ in accordance with that numerically observed at $Re = 1000$ by \cite{kashyap2020a} in large domains.
\section{Results}\label{sec:results}
\begin{figure}
	    \centering
        \subfigure[]{\includegraphics[trim=3cm 5cm 6cm 5cm,width=.425\columnwidth]{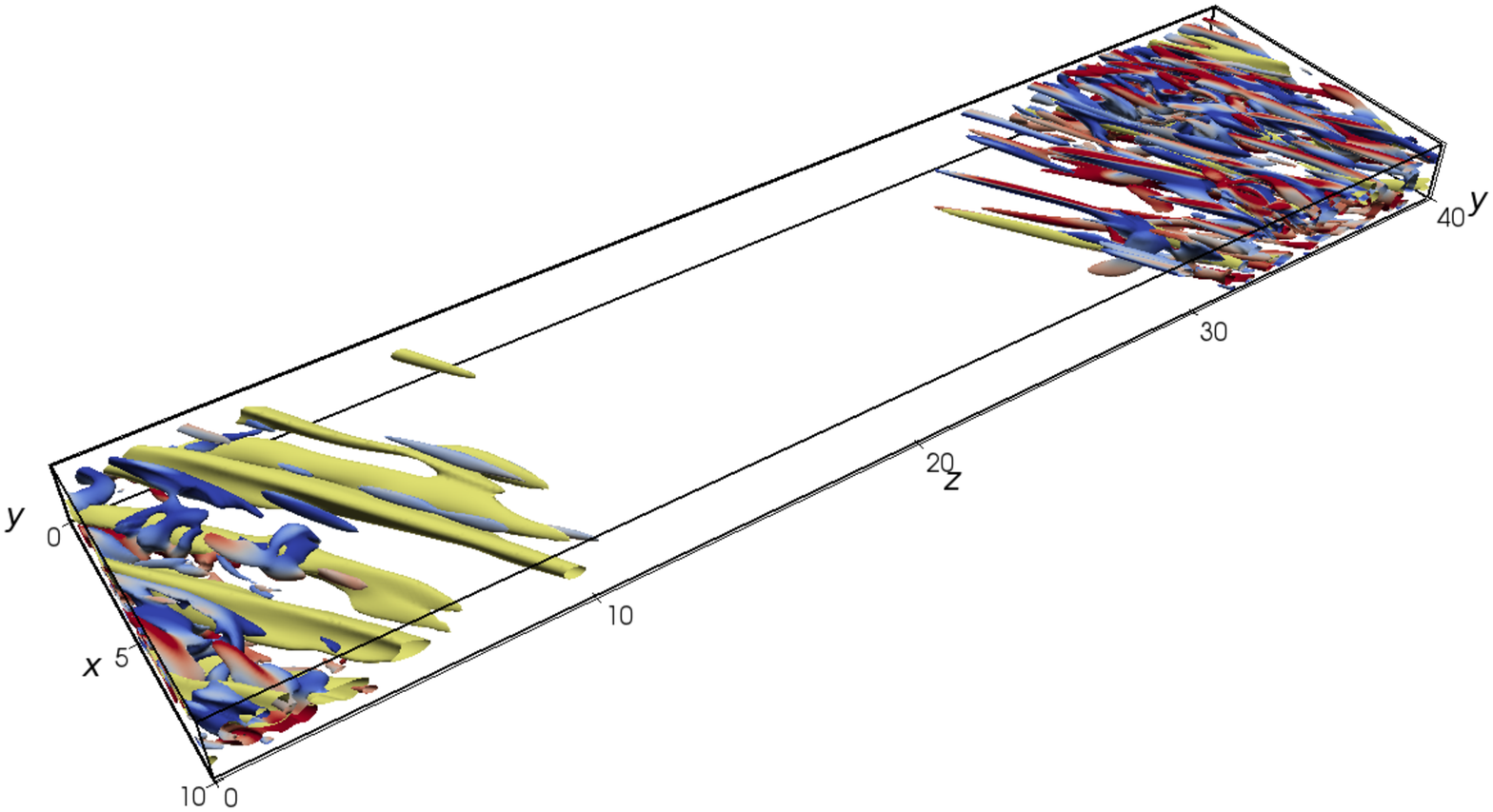}\label{snapshot}}
        \subfigure[$k_x E_{uu}(k_x)$]{\includegraphics[width=.28\columnwidth]{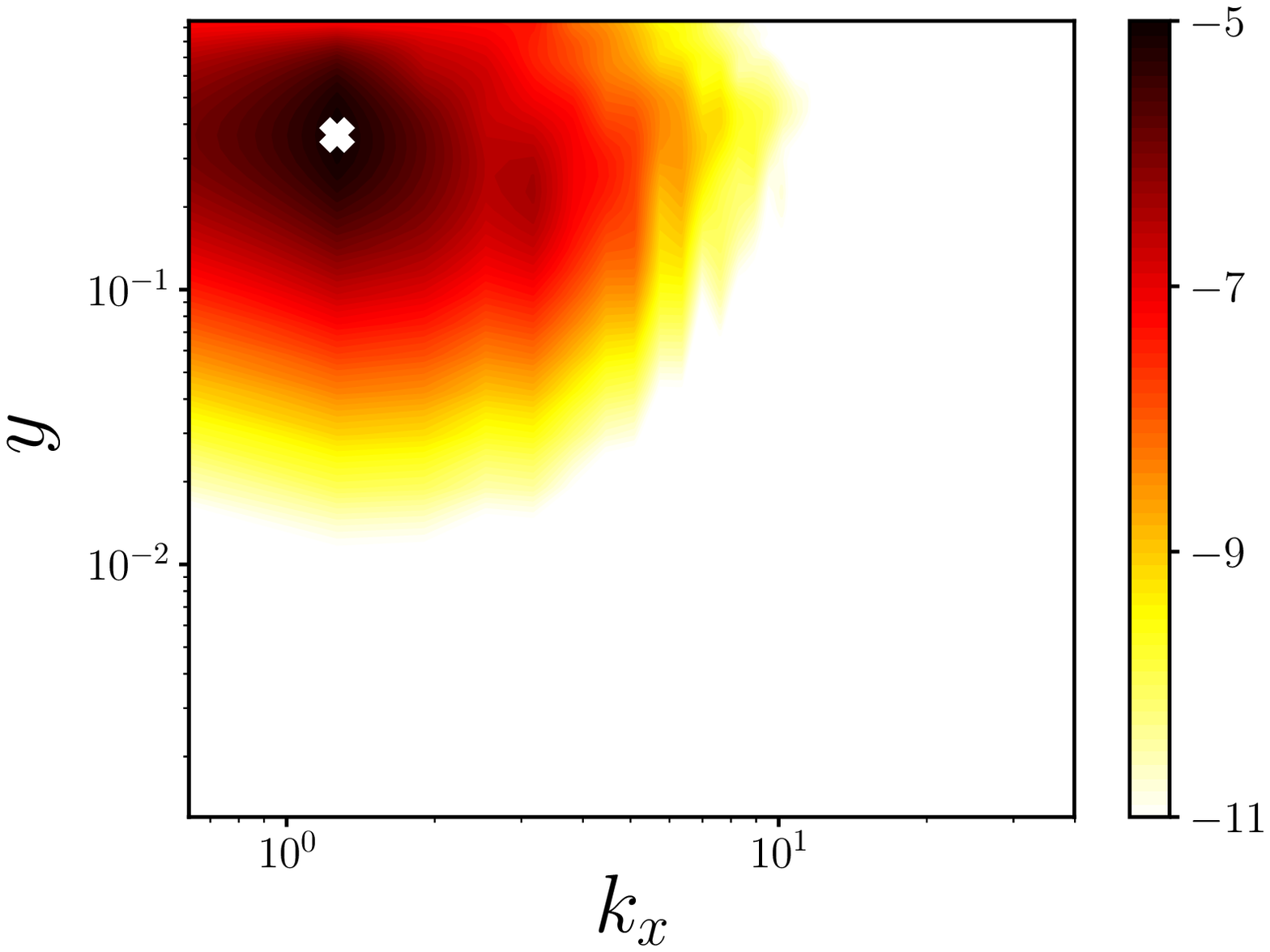}\label{spectrax}}
        \subfigure[$k_z E_{uu}(k_z)$]{\includegraphics[width=.28\columnwidth]{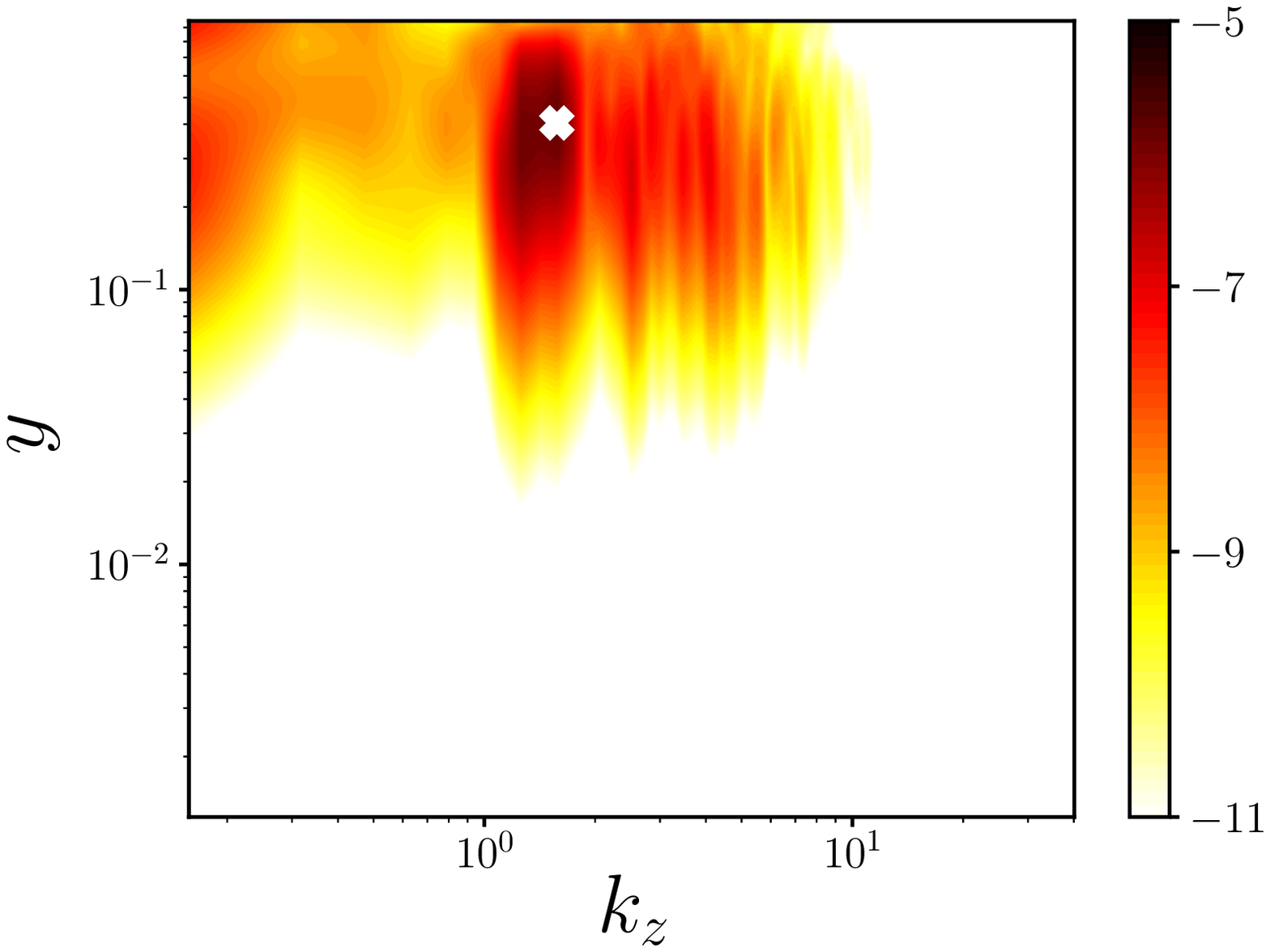}\label{spectraz}}
        \caption[]{(a) Isosurface of negative streamwise velocity ($u = -0.16$, yellow) and Q-criterion ($Q = 0.05$) coloured by the streamwise vorticity (positive red, negative blue)  of a turbulent-laminar pattern at $Re=1000$ in a domain tilted with $\theta=35^{\circ}$. (b-c) Logarithm of the premultiplied spectral energy versus the wall-normal distance $y$ for the instantaneous field in (a). The white 'X' symbols indicate the energy peaks.} 
        %ENZA: nella figura (a), si vede poco la localizzazione della banda, perché upstream e downstream si vedono molto gli streaks. Puoi cambiare il livello dei contorni in modo che si veda piu ''bianco'' al centro e che la banda risulti piu localizzata?
        \label{DNS}
\end{figure}
\begin{figure}
        \centering
        \subfigure[$t=0$]{\includegraphics[width=.4\columnwidth]{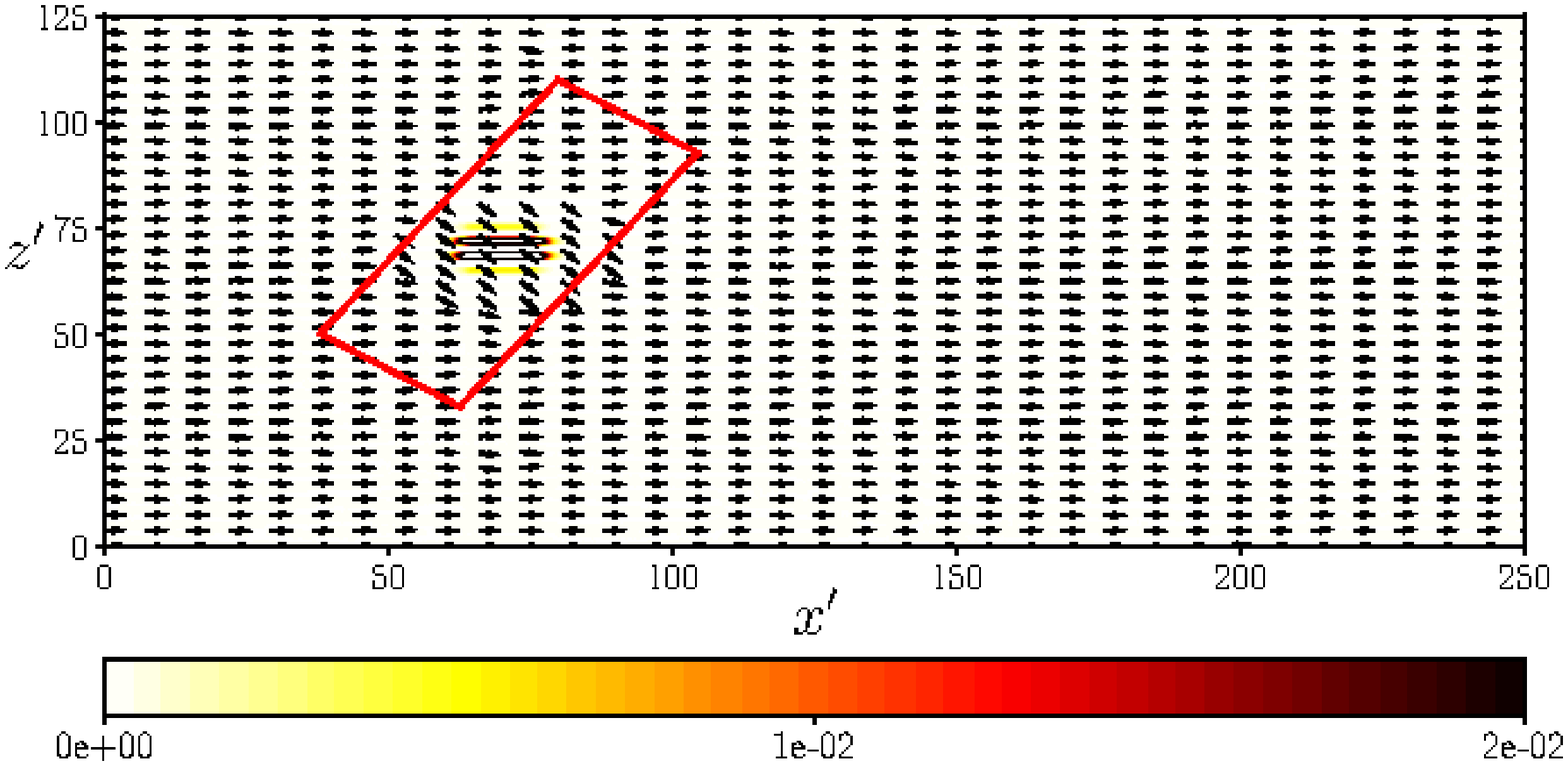}}
        \subfigure[$t=100$]{\includegraphics[width=.4\columnwidth]{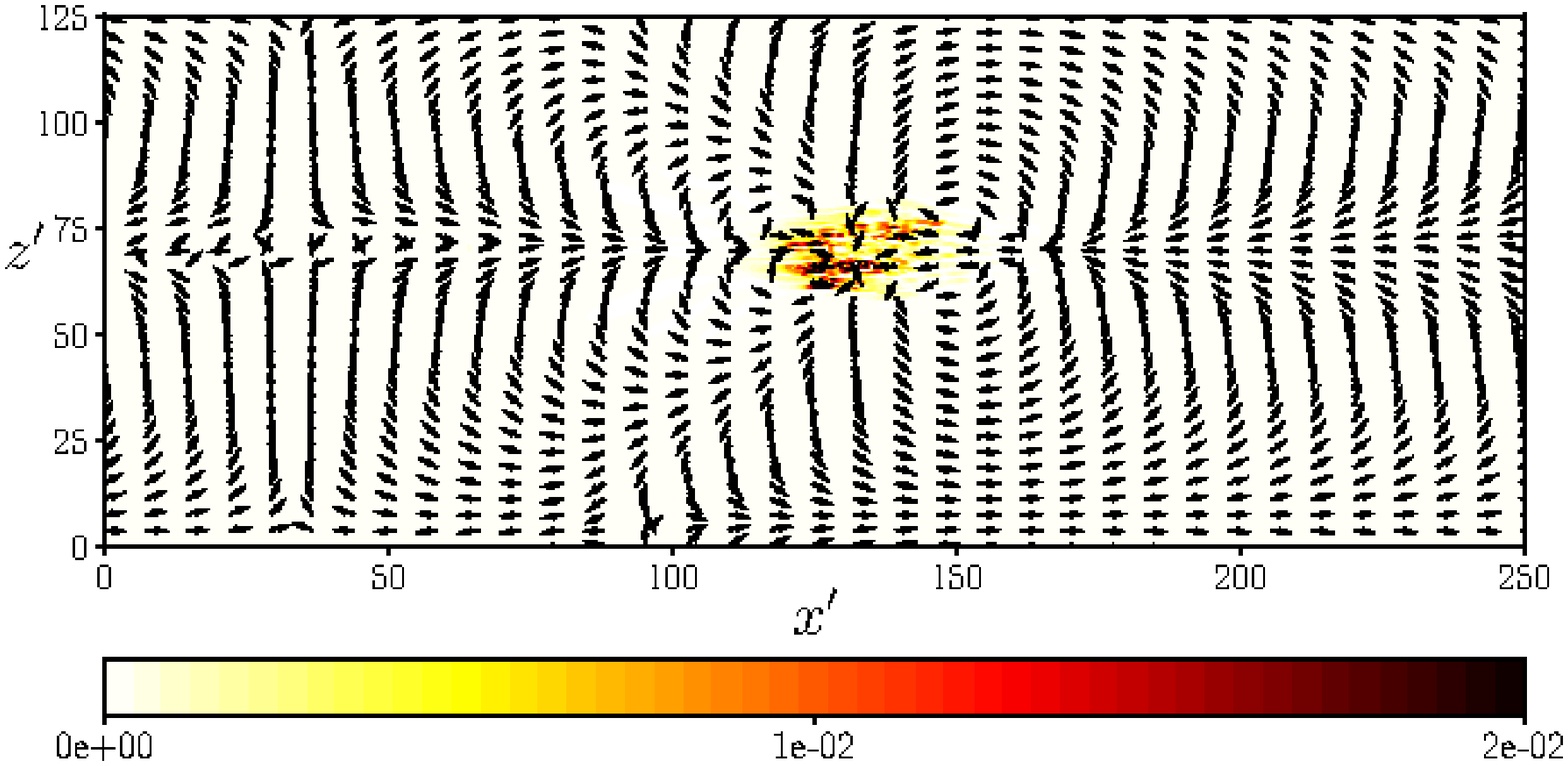}}
        \subfigure[$t=300$]{\includegraphics[width=.4\columnwidth]{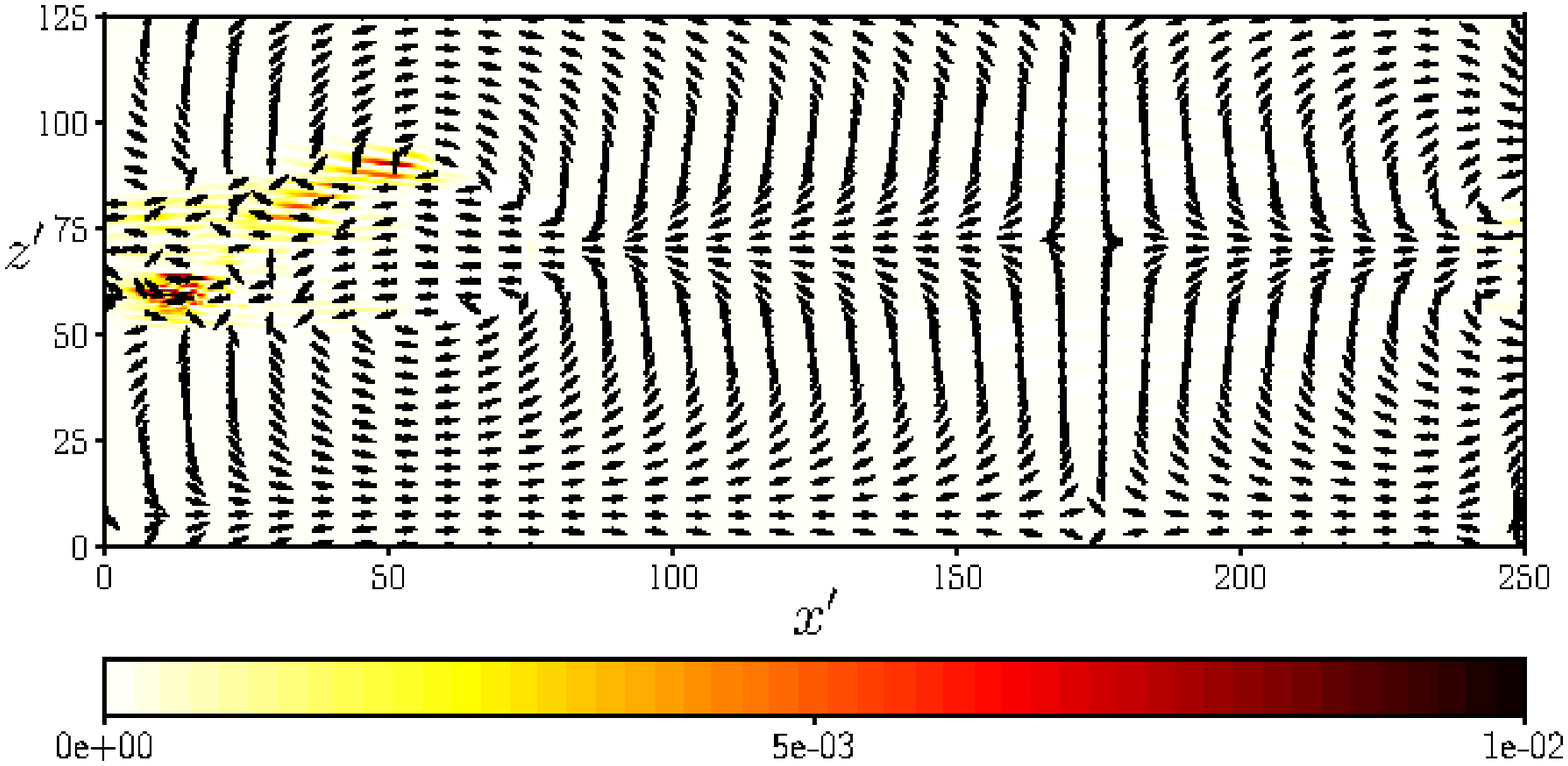}}
        \subfigure[$t=800$]{\includegraphics[width=.4\columnwidth]{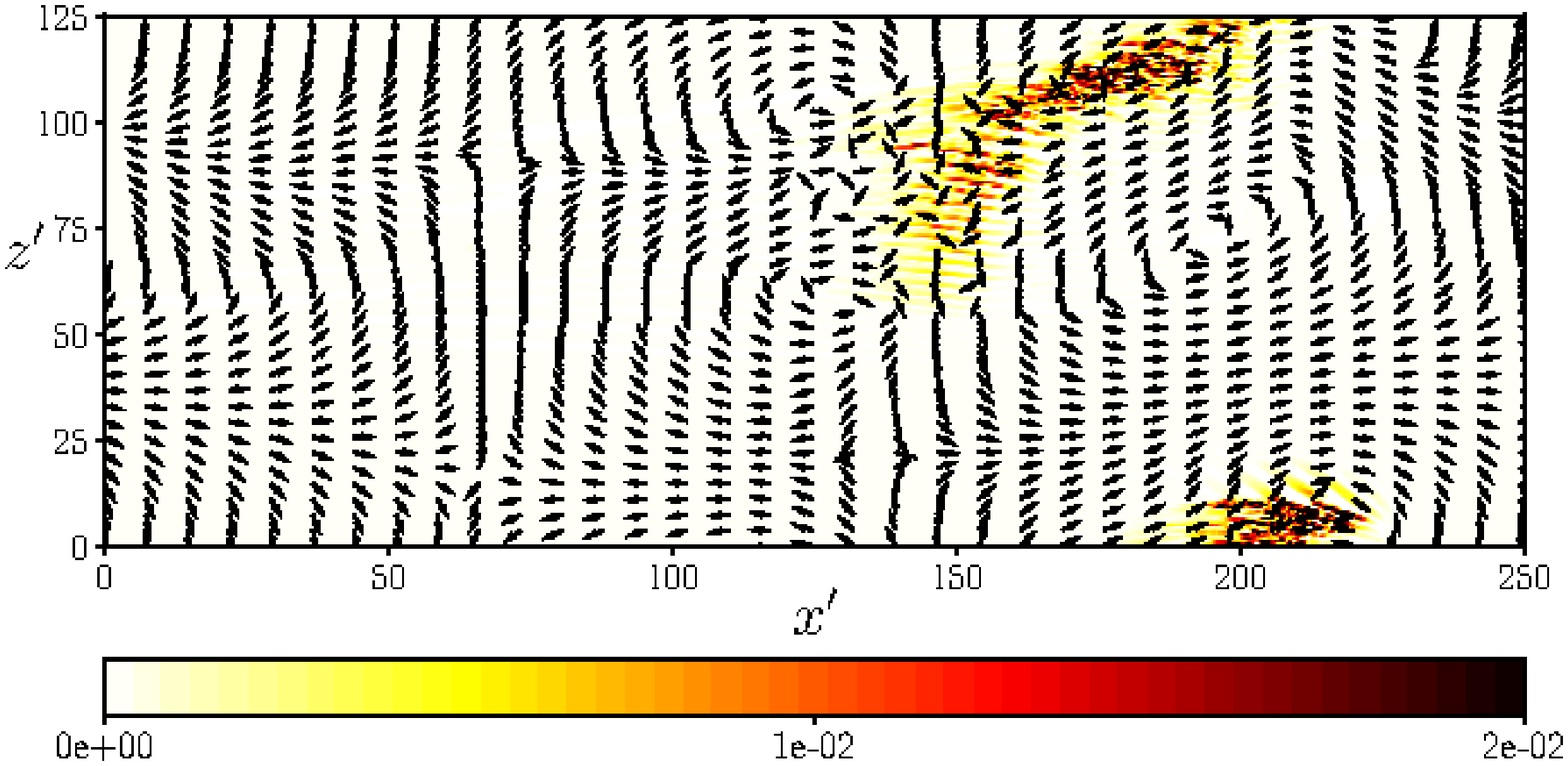}}
        \caption[]{Time evolution of the localised linear optimal perturbation reported in a large domain: contours of the wall-normal velocity and vectors of the large-scale flow ($\overline{u},\overline{w})$.} 
        \label{locLOPevolution-large}
\end{figure}

At first, a Direct Numerical Simulation (DNS) is performed at $Re = 1000$ in the tilted domain. In figure \ref{snapshot}, a snapshot of the perturbation field is shown. As previously done by other authors, the DNS is initialised with a Reynolds number for which turbulence occupies the whole domain. Then, the Reynolds number is reduced slowly until $Re = 1000$, reaching the laminar-turbulent pattern shown in figure \ref{snapshot}. As already discussed by \cite{tuckerman2014} for a tilted domain with the same size and Reynolds number, the turbulent state appears in the form of one turbulent band. % aligned with the streamwise direction.
In the instantaneous field, oblique wave-like structures such as alternating low- and high-speed streaks, are observed within the turbulent band and at its head. As expected, these structures present an angle with respect to the streamwise direction comparable to that of the base flow, and resemble the streaks observed at the head of a turbulent band in large (non-tilted) domains \citep{shimizu2019,liu2020}. Inspecting the premultiplied energy spectra of the streamwise instantaneous velocity provided in figure \ref{spectrax}, \ref{spectraz}, we found an energy peak at $k_x \approx \pm 1.27$, $k_z \approx \pm 1.6$. %In accordance with what mentioned above, 
Thus, as discussed above, the flow is dominated by oblique structures with angle $ \approx \arctan(k_x/k_z) \approx \pm 38^{\circ}$.\\
In order to understand the origin of these oblique structures and the main mechanisms responsible for the generation of a turbulent band, a linear optimization of perturbations  in the tilted domain is first performed. % to check if a linear mechanism can be involved. Thus, a
Since the base flow varies only in the wall-normal direction, the kinetic energy of the perturbations, constrained by equations \eqref{NS} linearized with respect to the base flow, is optimized using a local approach, where the perturbation is supposed to be sinusoidal in the streamwise and spanwise direction, with given wavenumbers $k_x$ and $k_z$, respectively.
\begin{figure}
        \begin{center}
        \subfigure[]{\includegraphics[width=.32\columnwidth]{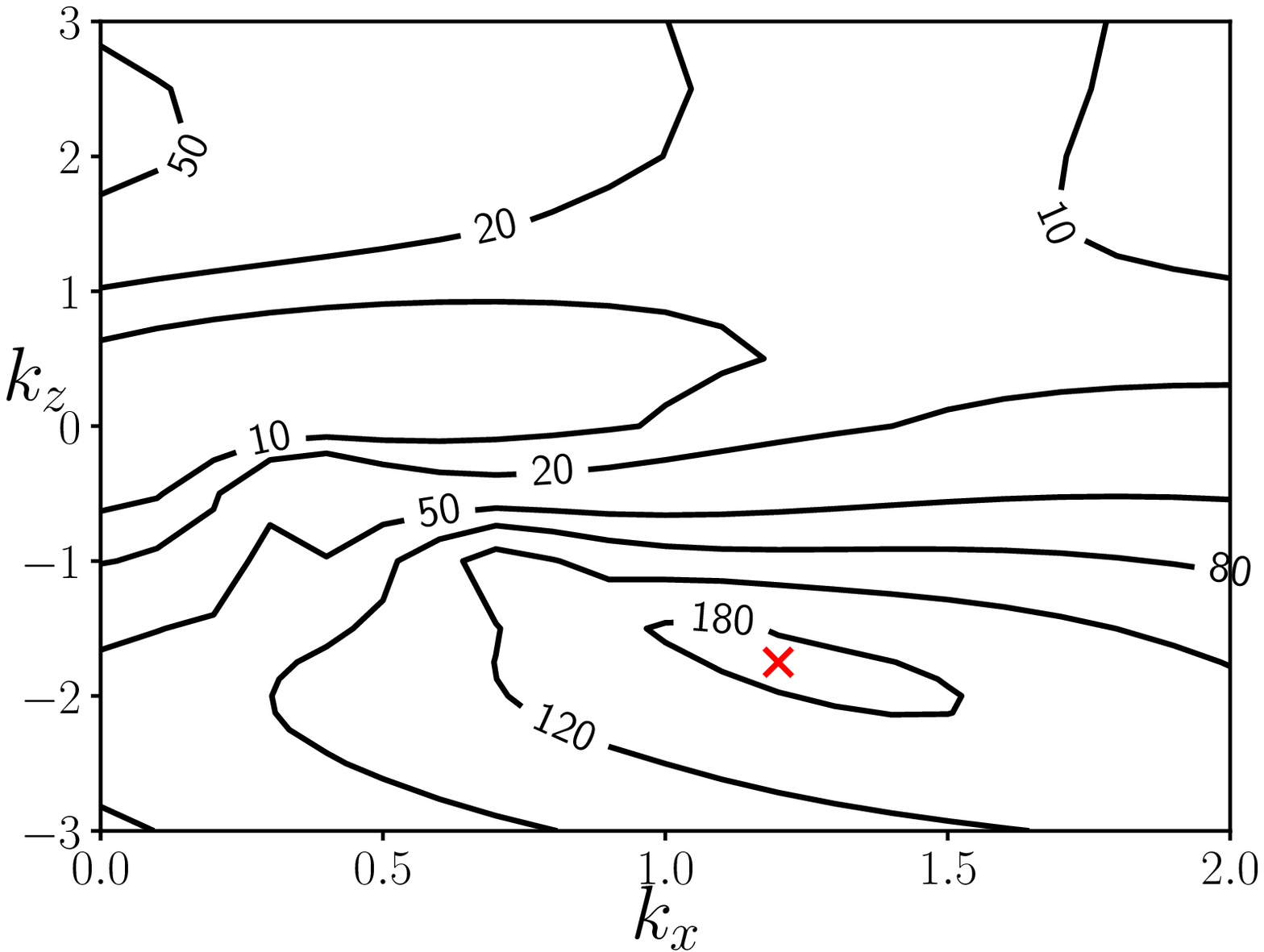}\label{LOP_map}}
     %   \subfigure[]{\includegraphics[width=.28\columnwidth]{./Figure_Re1000_angle35/map_Topt_alpha_beta_LOP_Re1000_angle35.eps}\label{LOP_map}}
        \subfigure[]{\includegraphics[width=.32\columnwidth]{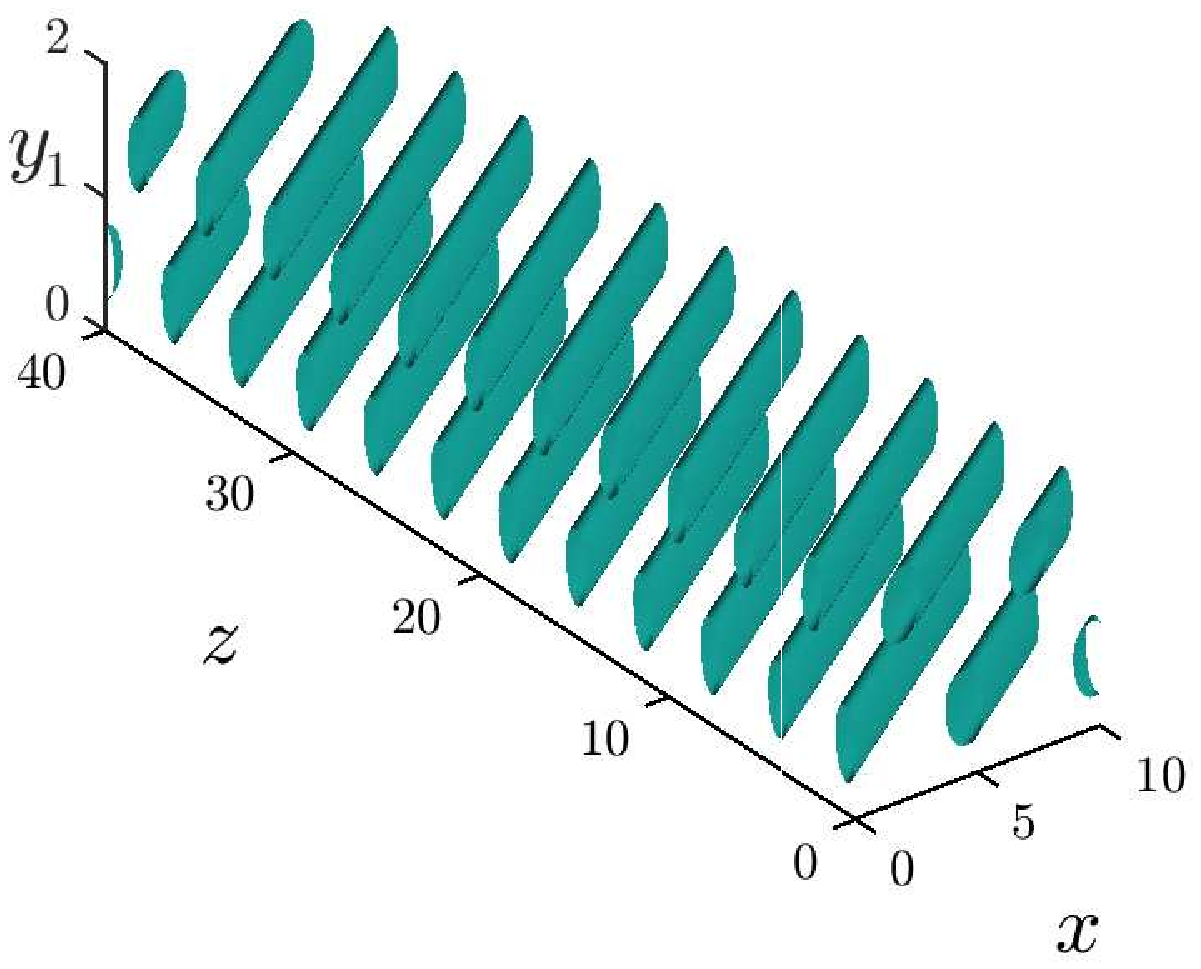} \label{3Dt0_LOP}}
        \end{center}
              \caption[]{(a) Contours of the optimal gain $G$ as a function of the streamwise ($k_x$) and spanwise ($k_z$) wavenumbers, for $Re=1000$ in the domain tilted with  angle $\theta = 35^{\circ}$. The red cross indicates the optimal growth $G_{opt}$. 
        %ENZA: aggiungere la red cross
        (b) Streamwise velocity component of the initial optimal perturbation
    %(c-d) Wall-normal profiles of the optimal disturbance at (c) $t=0$ and at (d) the optimal time
        for $T= T_{opt} = 73.11$, $k_x = 1.2$, 
    $k_z = -1.75$.} 
\end{figure}
\begin{figure}
        \centering
            \subfigure[Linear, $t=0$]{\includegraphics[width=.24\columnwidth]{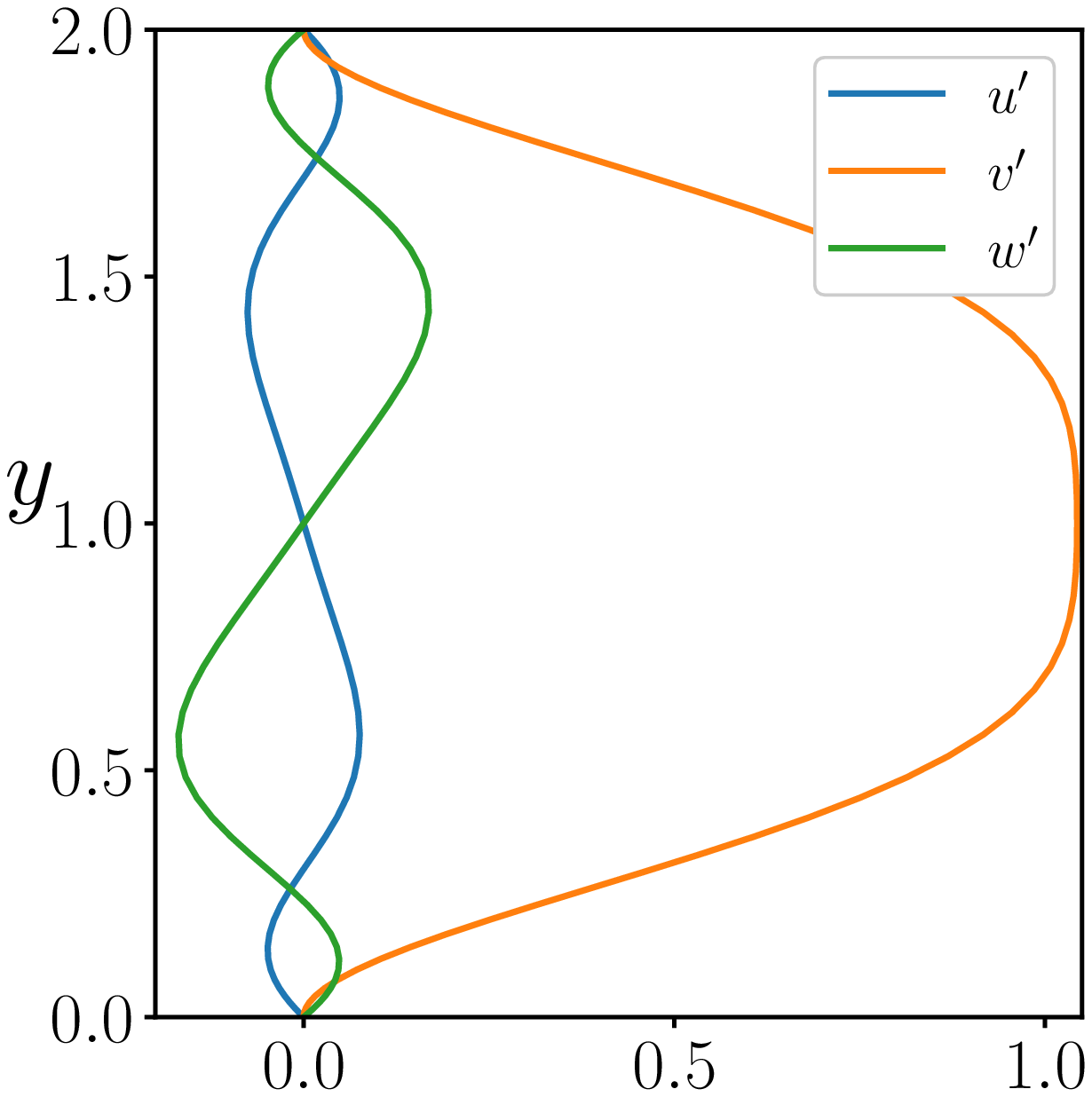} \label{1Dt0_LOP}}
        \subfigure[Linear, $t=T_{opt}$]{\includegraphics[width=.24\columnwidth]{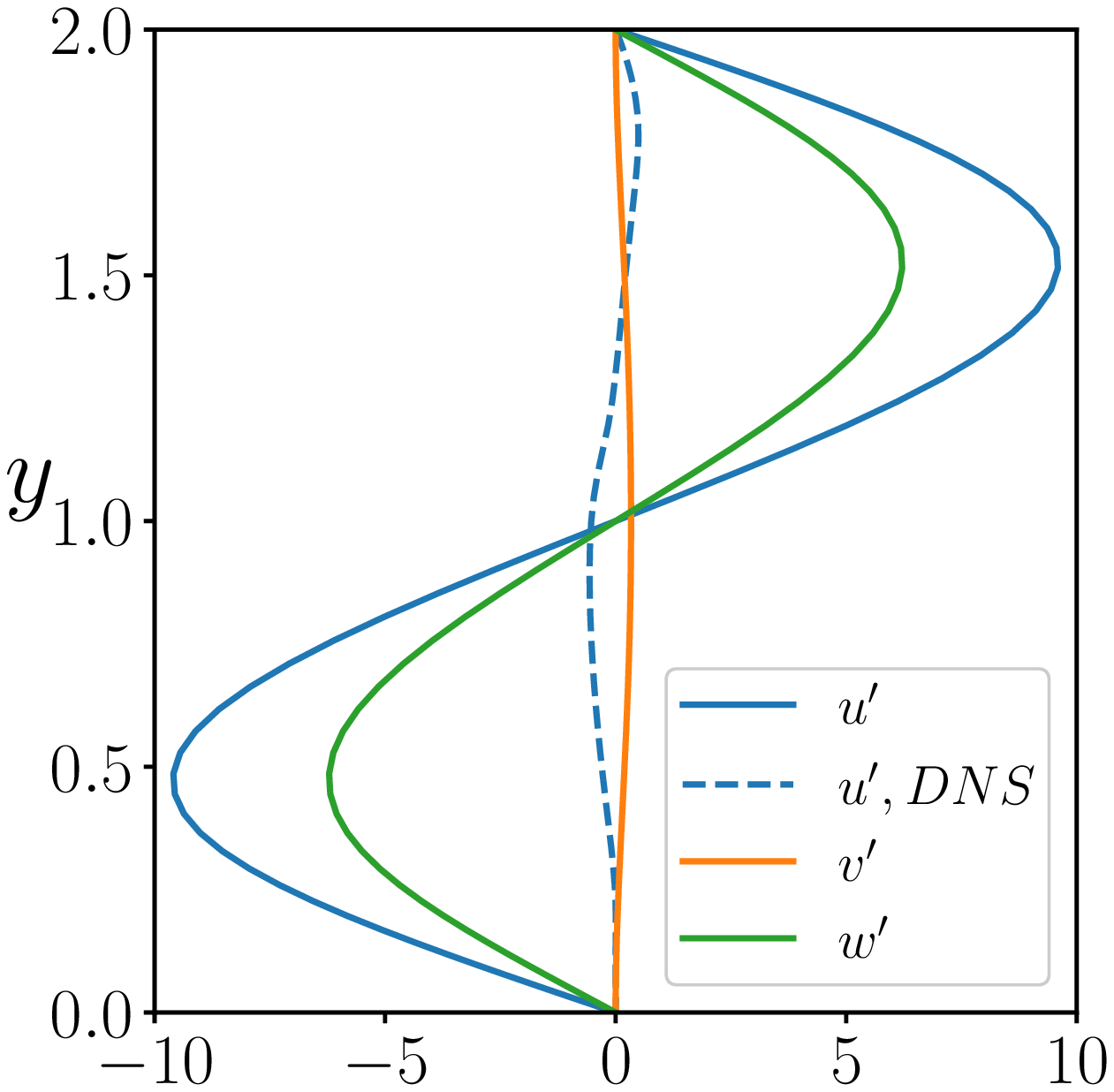}\label{1DtT_LOP}}
              \subfigure[Nonlinear, $t=0$]{\includegraphics[width=.24\columnwidth]{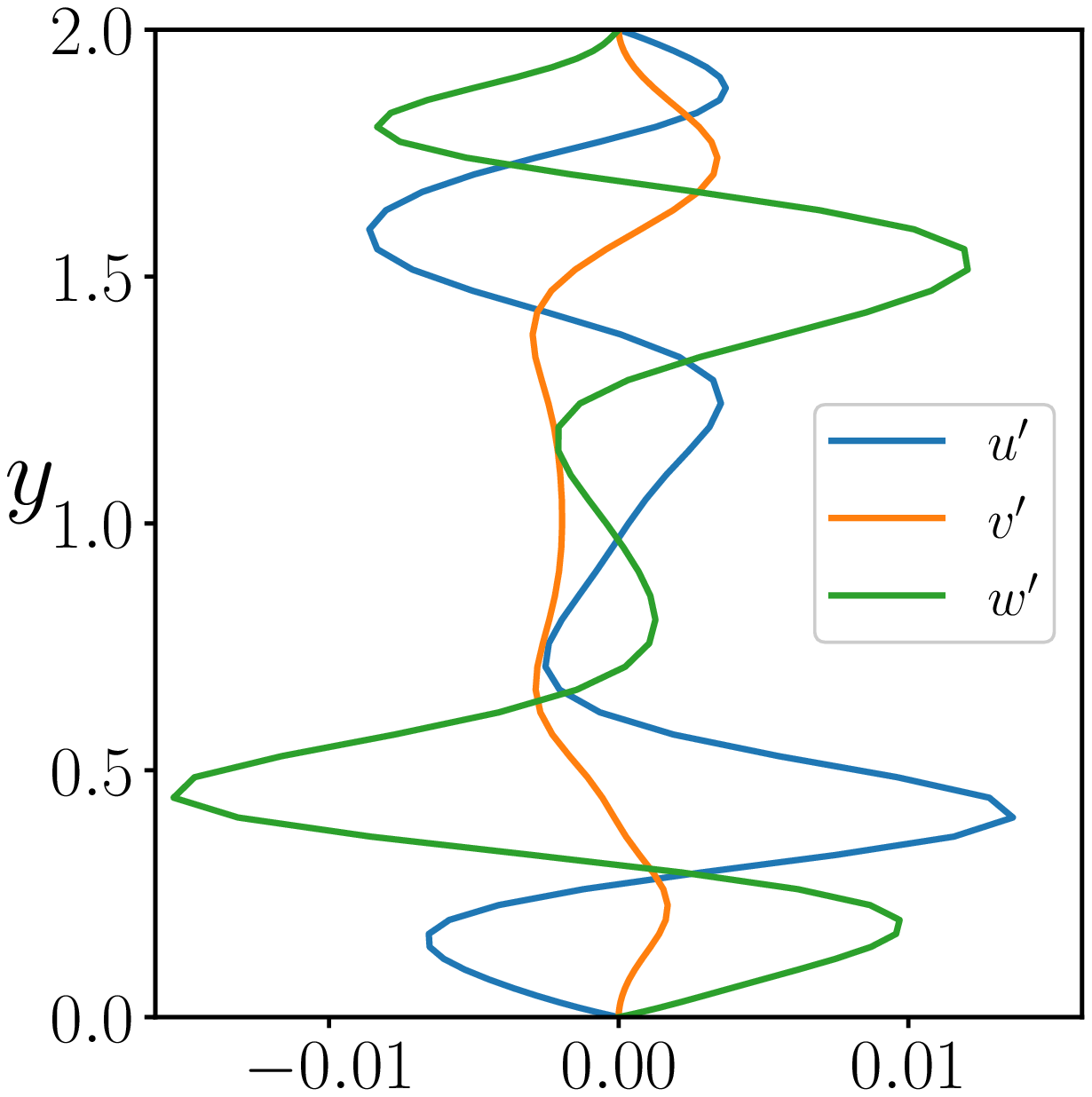}\label{1Dt0_NLOP}}
        \subfigure[Nonlinear, $t=T$]{\includegraphics[width=.24\columnwidth]{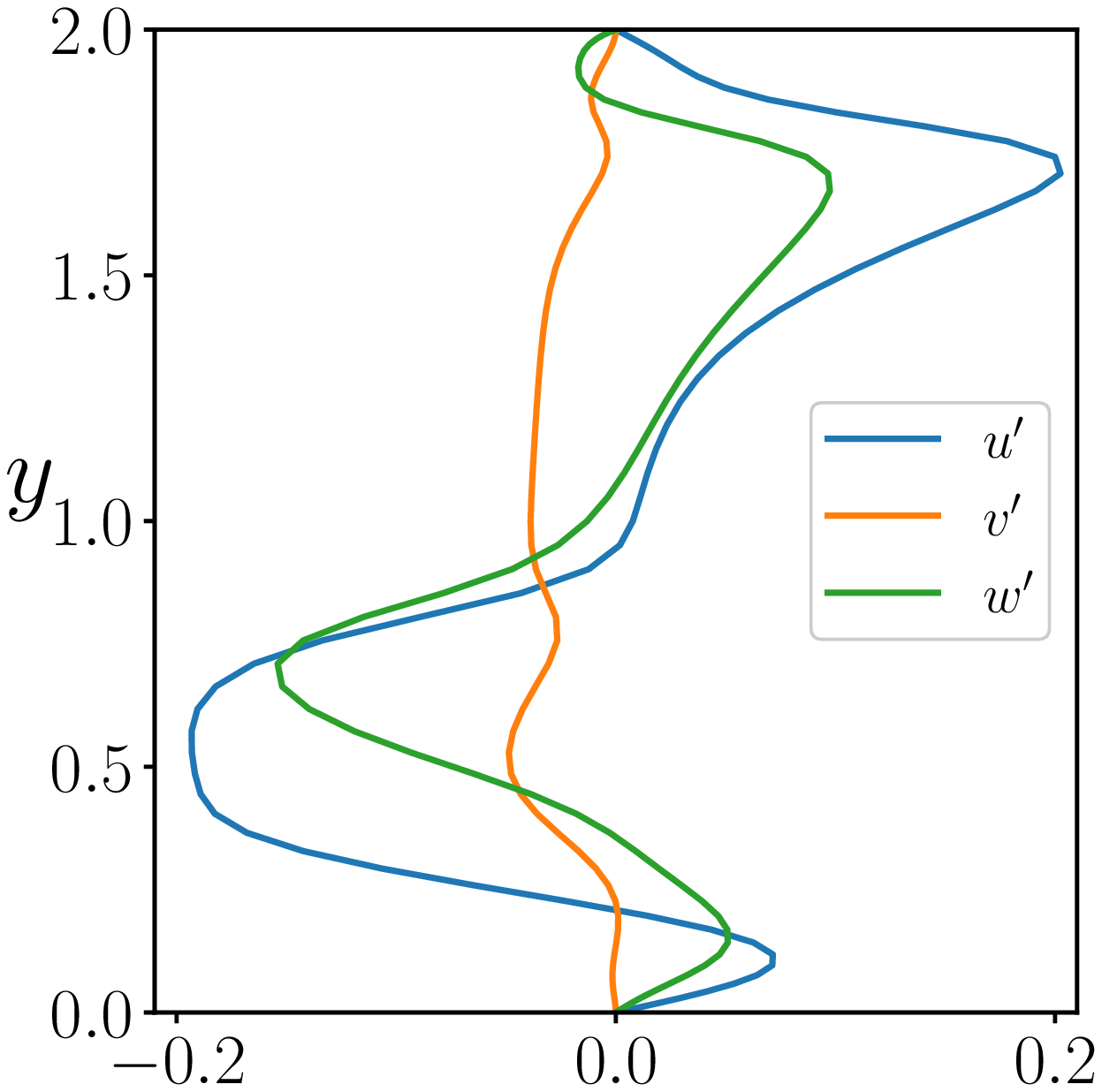}\label{1DtT_NLOP}}
      %  \subfigure[LOP in DNS, $t=5$]{\includegraphics[width=.35\columnwidth]{./Figure_Re1000_angle35/LOP_in_DNS_1Dplot_t5_x0_z10.eps}}
        \caption[Nonlinear, $t=T$]{ Velocity profiles of the optimal disturbances (continuous lines). (a-b) Linear optimal for $k_x = 1.2$, 
    $k_z = -1.75$  at (a) $t=0$ and (b) $T_{opt} = 73.11$. The dashed line represents the streamwise velocity profile recovered at $x=0$, $z=10$ by evolving to $t=5$ by DNS the localized initial optimal perturbation of figure \ref{locLOPevolution}.  (c-d) Nonlinear optimal at the minimal energy able to trigger turbulence, $E_0=2.1\times 10^{-5}$ at (c) $t=0$ and (b) $T = 10$.} 
\end{figure}
The linear optimization problem was solved  at $Re = 1000$ for streamwise and spanwise wavenumbers in the range $0<k_x<2$, $-3<k_z<3$. In figure \ref{LOP_map}, is provided the variation of the optimal gain with the spanwise and streamwise wavenumbers. The maximum growth is achieved at the optimal target time $T_{opt} = 73.11$, for $k_x = 1.2$ and $k_z = -1.75$, leading to an optimal gain  $G_{opt}=195.28$. It is interesting to note that the optimal gain and time values are very similar to those found by \cite{reddy1993} for the plane Poiseuille flow, although obtained for different wavenumbers. Whereas, similar values of streamwise and spanwise wavenumbers are found by \cite{xiao2020} performing a linear stability analysis around the mean flow of a region at the head of the turbulent band. 
Moreover, the optimal streamwise and spanwise wavenumbers are very close to the ones for which the premultiplied energy spectra in figure \ref{spectrax}, \ref{spectraz} peak. Thus, these optimal perturbations can be linked to the oblique waves observed at the head of the turbulent band.
As shown in figure \ref{3Dt0_LOP}, the linear optimal perturbation is oblique with angle  $\arctan(k_x/k_z) \approx -34.5^{\circ}$,  %, in accordance to the classical plane Poiseuille flow 
%\citep{reddy1993} 
and modulated in both streamwise and spanwise directions. This had to be expected since the base flow presents a spanwise component, in analogy with the shear flow developing on a swept-wing, whose unstable modes and optimal perturbations are characterised by cross-flow vortices, namely three dimensional oblique vortical perturbations with negative spanwise wavenumber. As shown in figure \ref{1Dt0_LOP}, at $t=0$ the optimal perturbation presents counter-rotating vortices with a large wall-normal component, which decreases in time towards the target time (see figure \ref{1DtT_LOP}), while the streamwise and spanwise ones strongly increase creating oblique streaks. The mechanism creating these oblique energetic structures is based on the transport of the wall-normal shear of both streamwise and spanwise component of the base flow, which may be seen as a tilted counterpart of the lift-up effect.
%Thus, at target time the optimal perturbation is characterized by oblique streaks tilted with respect to the streamwise direction with an angle of $arctg(k_x/k_z) \approx 34.5^{\circ}$ (see figure \ref{3Dt0_LOP}).
\\
%Looking at the shape of the optimal perturbation along the wall-normal direction in figure (\ref{1Dt0_LOP},\ref{1DtT_LOP}), it seems that the mechanism involved in the transition to turbulence is the classical lift-up mechanism. In fact, the initial optimal perturbation in figure \ref{1Dt0_LOP} is characterised by  given a wall-normal velocity bigger than the other velocities components $u-w$. And as typically occurs in the case of the lift-up mechanism at the optimal time $t=t_{opt}=73.11$ (see figure \ref{1DtT_LOP}), the wall-normal velocity components $v$ is small compared with the streamwise velocity. Moreover, due to the spanwise modulation of the optimal perturbation, the spanwise velocity component is comparable to the streamwise velocity. 
%This results in a linear optimal perturbation characterised by oblique streaks tilted with respect to the streamwise direction with an angle of $arctg(k_x/k_z) \approx 34.5^{\circ}$, that implies that the streaks are aligned with the base flow (see figure \ref{3Dt0_LOP}).\\
\begin{figure}
	    \centering
       \includegraphics[width=.5\columnwidth]{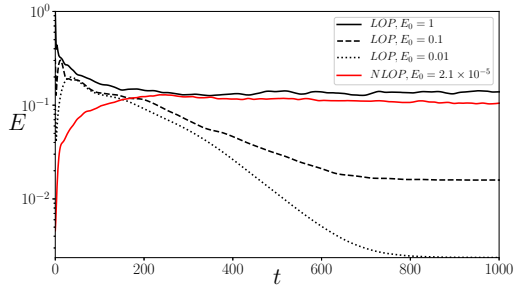}
        \caption[]{Kinetic energy time evolution for the  linear optimal perturbation with $E_0 = 0.01.0.1,1$ (black lines) and for the nonlinear one (red line) for $E_0=2.1\times 10^{-5}$. 
        %ENZA: puoi tagliare questa figura in modo che sia piu visibile la parte iniziale della crescita dell'energia, tipo plottando fino a t=1000? Perché cosi sembra quasi che l'energia scenda a tempo piccolo, e non salga proprio. Inoltre, io aggingerei un secondo subplot che faccia vedere l'energia nel tempo per l'ottimale non lineare. In entrambe le sottofigure, sull'asse delle ordinate invece di E_k metti E. 
        }
        \label{Ek_in_time}
\end{figure}
The linear optimal perturbation computed for $T_{opt}$ is then injected onto the laminar flow in the tilted domain with different values of the initial energy $E(0)$, in order to verify whether such a linear transient-growth mechanism could induce transition in the form of turbulent bands. In figure \ref{Ek_in_time}, the energy evolution in time is reported for the linear optimal perturbations with different initial energies (black lines). The perturbation with unitary energy norm is the only one able to induce the formation of the turbulent band, while the others lead to relaminarisation. However, it is observed that turbulence is at first triggered in the whole domain and successively localises in a band. %However, the other perturbations reach an intermediate state unable to trigger turbulence neither in a localised region nor in all the domain.\\
%Thus, it appears that the linear optimal perturbation is not well adapted for triggering a turbulent band.
%It is clear that turbulence can not be triggered with this linear optimal perturbation as it requests a huge energy level. Moreover, the linear optimal perturbation is not able to localise the flow. 
This is mostly probably due to the fact that the linear optimal disturbance is not spatially localized but occupies the whole domain, which also explains the large amount of energy needed for triggering turbulence by means of this optimal mechanism. To provide spatial localization of the optimal perturbation, aiming at triggering the turbulent band, we extended the optimization to the fully nonlinear equations, which usually provide a consistent spatial localization \citep{kerswellARFM,cherubini2010,Farano2015}. Notice that the nonlinear optimization is performed in a fully three-dimensional framework, without any hypothesis on the perturbation wavenumbers. \\
%Then, we can assume that maybe a nonlinear mechanism is able to localise the flow and to develop turbulent bands, for this reason a nonlinear optimization problem is formulated.\\ 
\begin{figure}
	    \centering
        \subfigure[2D, $t=0$ ]{\includegraphics[width=.14\columnwidth]{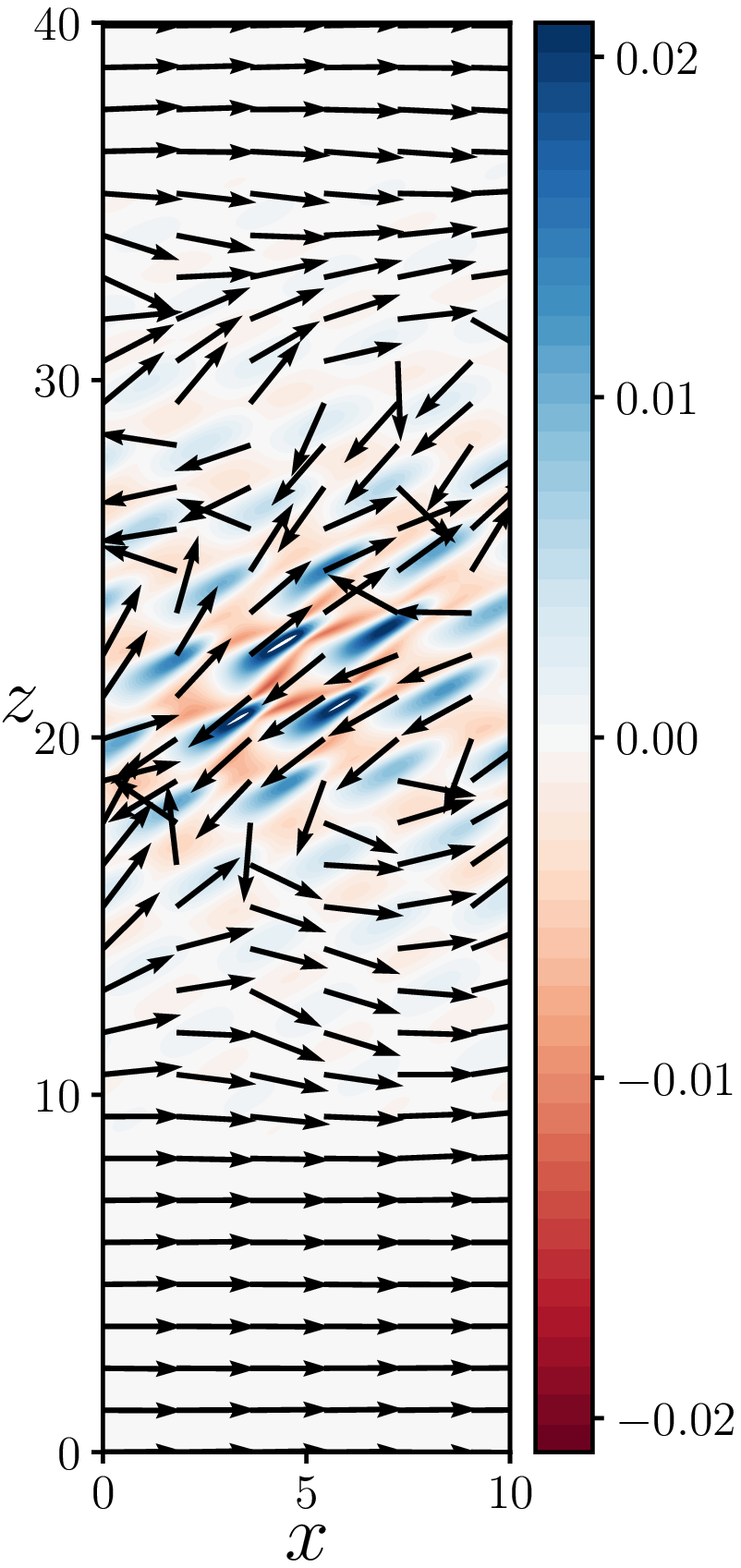}\label{flowscales_NLOP}}
        \subfigure[3D, $t=0$]{\includegraphics[width=.40\columnwidth]{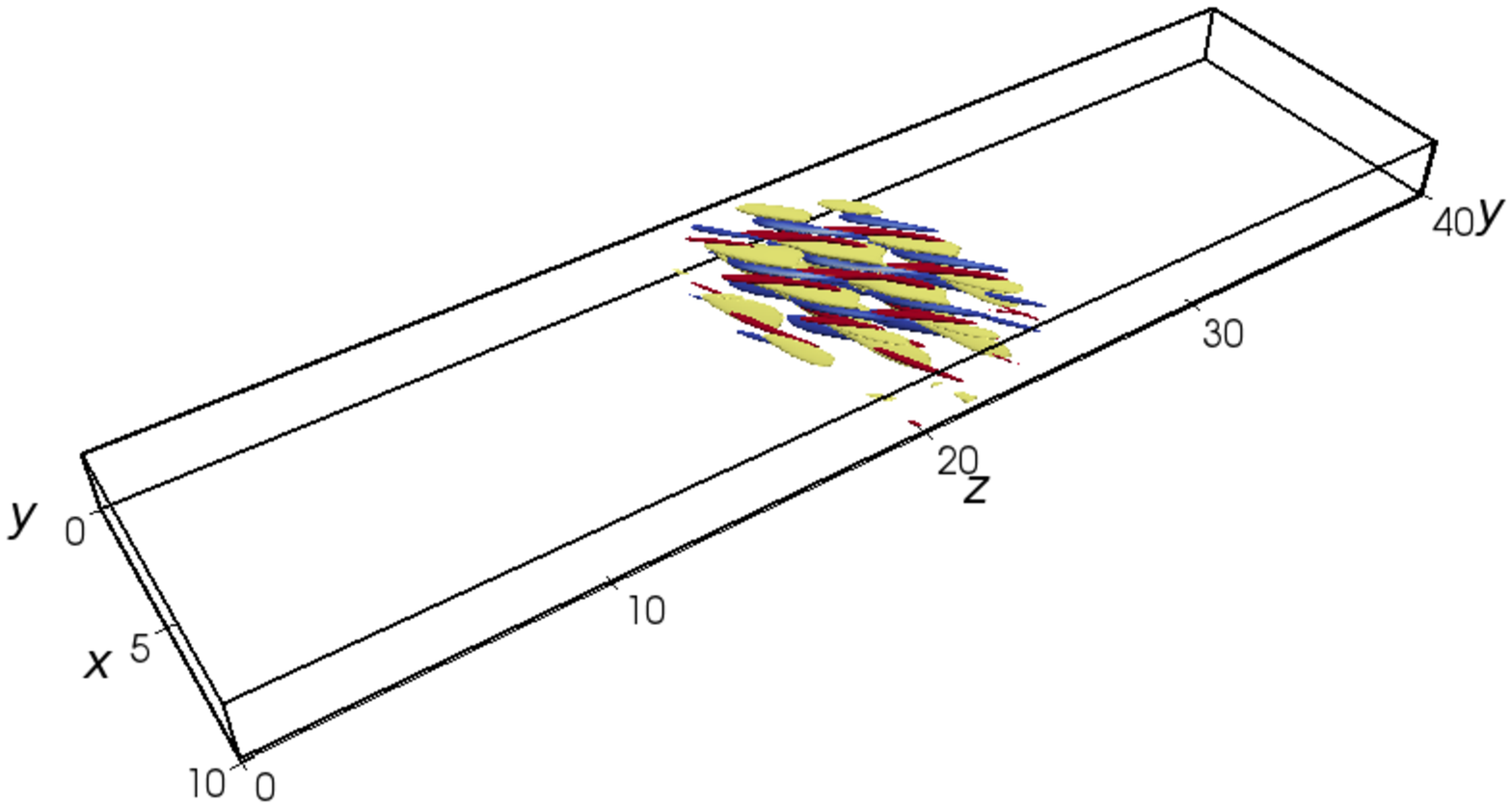}\label{3dflow_NLOP}}
        \subfigure[3D, $t=600$]{\includegraphics[width=.40\columnwidth]{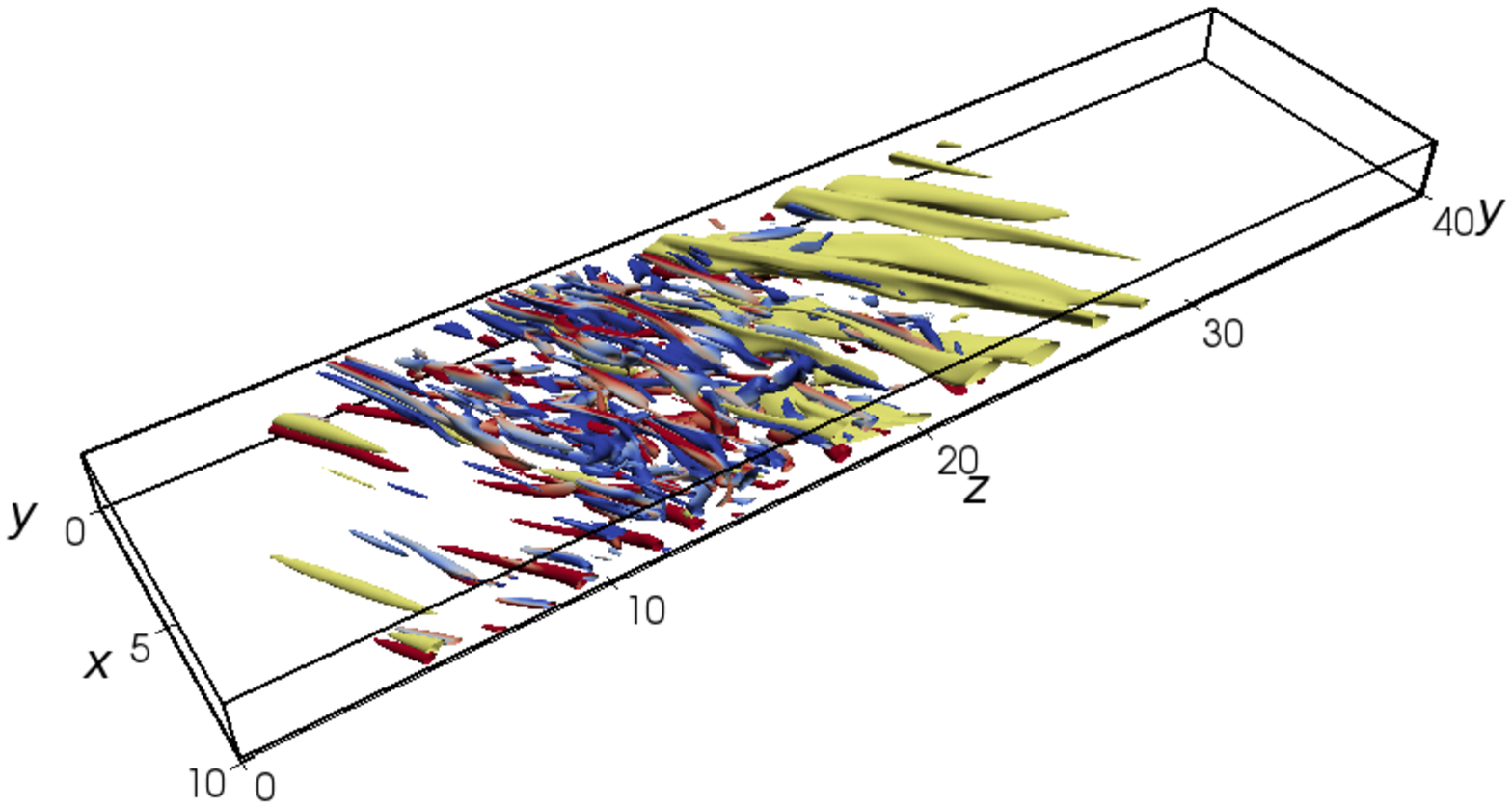}\label{3dflow_NLOP_T}}
        \caption[]{Nonlinear optimal perturbation for $Re = 1000$, $E_0 = 2.1 \times 10^{-5}$, $T = 10$. (a) Shaded isocontours of the wall-normal perturbation and vectors  of the $y$-integrated flow in the $y=0.25$ plane at $t=0$. (b-c) Isosurface of negative streamwise velocity (yellow) and Q-criterion coloured by the streamwise vorticity (positive red, negative blue) for (b) $t=0$, $u = -0.01$, $Q = 0.02$; (c) $t=600$, $u = -0.16$, $Q = 0.05$.}
        \label{NLOP}
        %ENZA: nella figura, modificare la cross-flow energy con la energy della componente v, oppure la componente v stessa. Poi nel testo modifico io. 
        %ENZA: secondo me qui, oltre alla perturbazione a tempo t=0, dovremmo far vedere anche la sua evoluzione in banda turbolenta, quindi ci aggiungerei anche un plot 3D a t piu alta. Inoltre, vorrei vedere dei profili rispetto a y, per capire se sono simili a quelli della perturbazione lineare.. Possiamo aggiungerli nella stessa figura, oppure metterne una nuova, lo spazio c'è. 
\end{figure}
 Nonlinear optimization has been performed in the tilted domain for several initial energies and target time $T=10$, which is close to the characteristic eddy turnover time of structures in the buffer layer, for which optimal streaks having the typical spanwise spacing of approximately $100$ wall units  were recovered by \cite{ButlerFarrell1993}. For this target time, the nonlinear optimal perturbation triggers localised turbulence already for $E_0 \ge 2.1 \times 10^{-5}$. The nonlinear optimal perturbation at the minimal input energy able to trigger turbulence is shown in figure \ref{NLOP} (a-b).  As expected from previous works \citep{cherubini2011,monokrousos2011,pringle2012}, it is localised in the longitudinal direction. Furthermore, it presents remarkable similarities with the edge state found by \cite{paranjape2020} in a tilted domain for $Re = 760$. %, by means of the similar meaning of the edge state and the nonlinear optimal solution at the minimal energy.
 %$$\\
%Moreover, it is characterised by two flow scales. 
In figure \ref{flowscales_NLOP} the isocontours of the wall-normal perturbation are reported, together with the normalised $y$-integrated large-scale flow $\overline{u_i} = \int_{-1}^1 u_i dy$. One can observe a small-scale flow within a localised region, where the turbulent band will be generated, together with two larger-scale vortices surrounding this region, having opposite direction upstream and downstream of the localised perturbation. A large-scale vortical flow surrounding the region developing into a turbulent spot has been previously reported by several authors in both tilted and non-tilted domains. The three-dimensional visualization in figure \ref{3dflow_NLOP} shows that the small-scale flow is constituted by oblique streaks flanked by counter-rotating vortices. The streaks are aligned with the base flow, presenting an angle of approximately $35^{\circ}$ with respect to the streamwise direction, in accordance with the angle of the linear optimal perturbation.
As expected, this localised optimal perturbation evolves in time towards a turbulent band, as shown in figure \ref{3dflow_NLOP_T}. Notice that the nonlinear optimal perturbation induces transition for an initial energy five orders of magnitude lower than that of the linear optimal one; this cannot be exclusively due to its spatial localisation. In fact, the wall-normal velocity profiles provided in figure \ref{1Dt0_NLOP} present strong differences with respect to their linear counterpart shown in figure  \ref{1Dt0_LOP}. In particular, as typically observed in nonlinear optimal perturbations \citep{cherubini2011}, the streamwise velocity component is now of the same order of magnitude than the other ones, and the wall-normal component strongly changes. At target time (figure  \ref{1DtT_NLOP}) deformed streaks are obtained, presenting inflection points which might be linked to the inflectional instability discussed in \cite{song2020}.\\
%As previously supposed, the turbulent band localisation is linked to a nonlinear effect. Then, it remains to be seen if the small scale flow can be linked to a linear mechanism. 
To isolate the effect of spatial localisation from the strong changes in the velocity profiles induced by the nonlinear effects, we enforced localisation in the $z$ direction on the three-dimensional linear optimal solution shown in figure \ref{3Dt0_LOP}. This is achieved by multiplying the velocity components for a normal distribution having the form: 
\begin{equation*}
    \textit{f}(z) = e^{-\frac{1}{2}\frac{(z-z_0)^2}{\sigma^2}},
\end{equation*}
where $z_0 = 10$ represents the value at which the perturbation should be centered, and $\sigma = 2.5$ is its standard deviation. %In this way, the optimal perturbation presents a shape very similar to the nonlinear optimal perturbation.\\
\begin{figure}
        \centering
        \subfigure[$t=0$]{\includegraphics[width=.15\columnwidth]{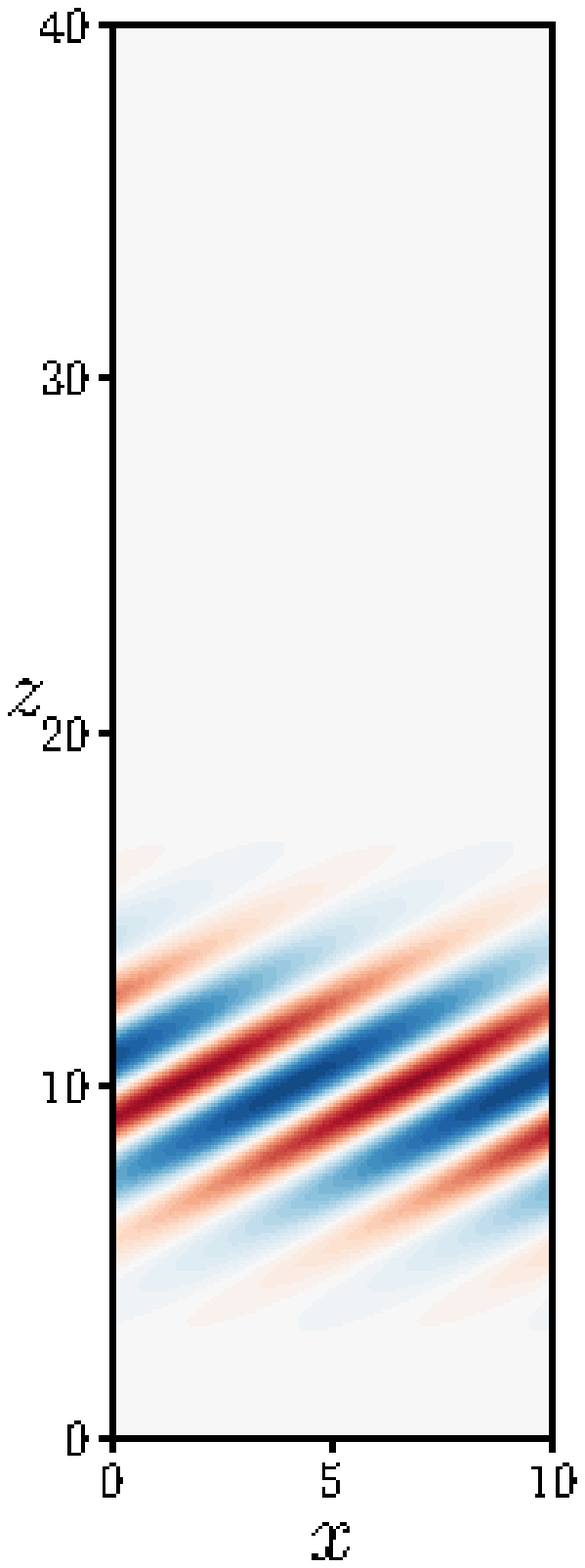}}
        \subfigure[$t=20$]{\includegraphics[width=.15\columnwidth]{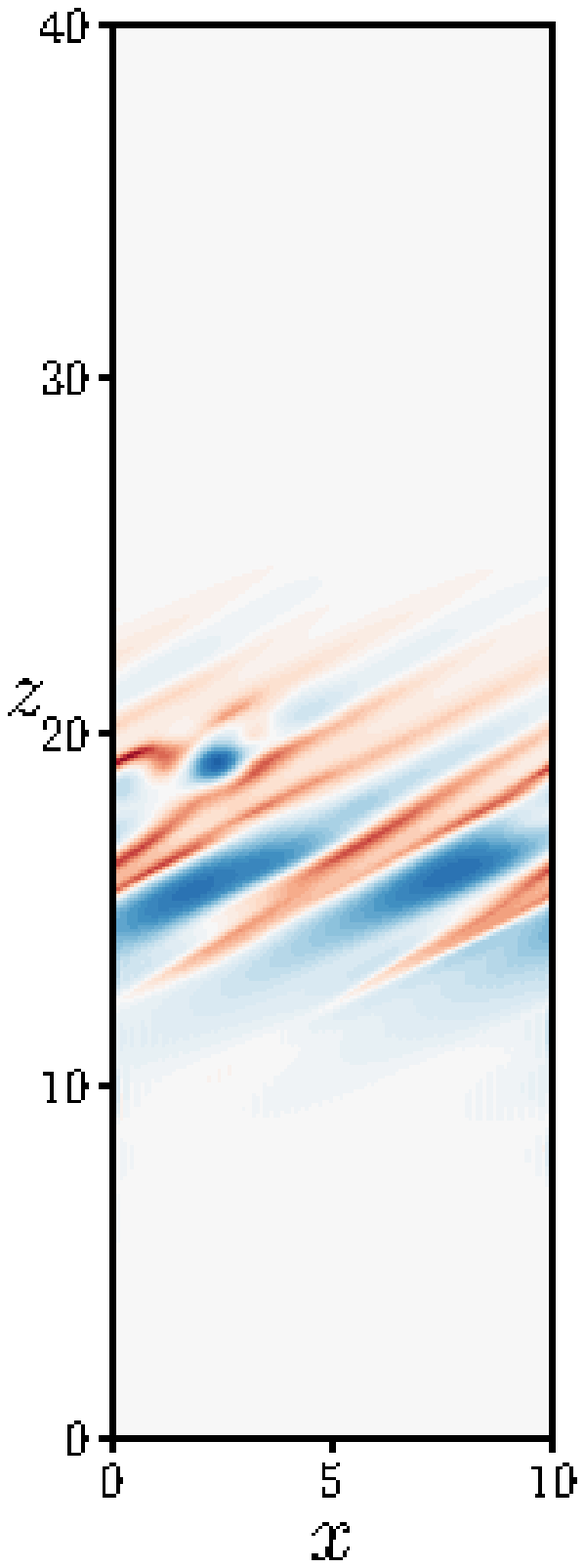}}
        \subfigure[$t=50$]{\includegraphics[width=.15\columnwidth]{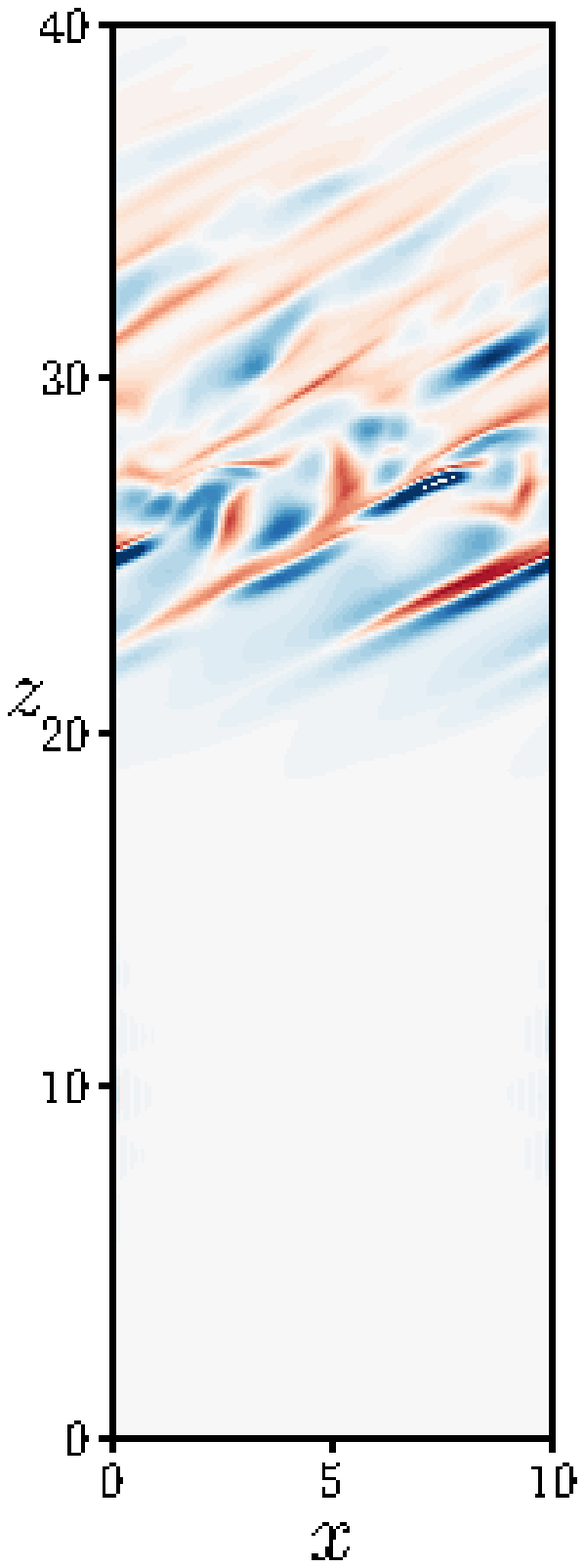}}
        \subfigure[$t=80$]{\includegraphics[width=.15\columnwidth]{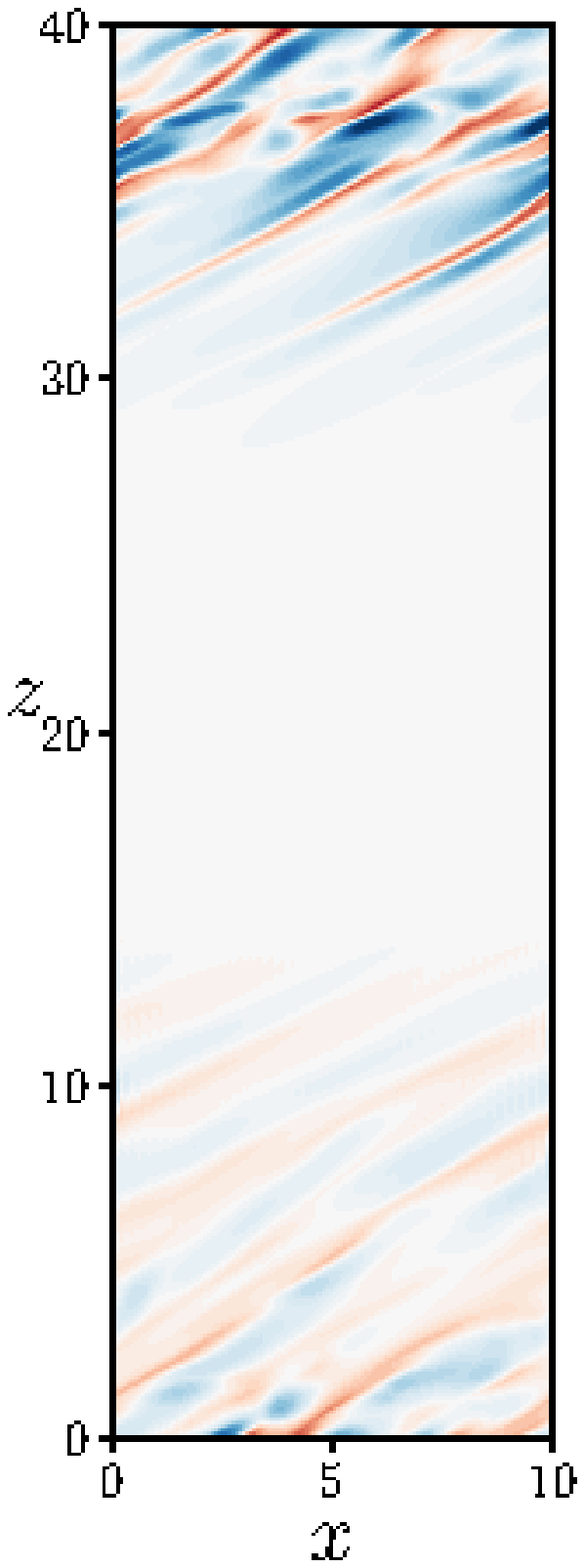}}
        \subfigure[$t=100$]{\includegraphics[width=.15\columnwidth]{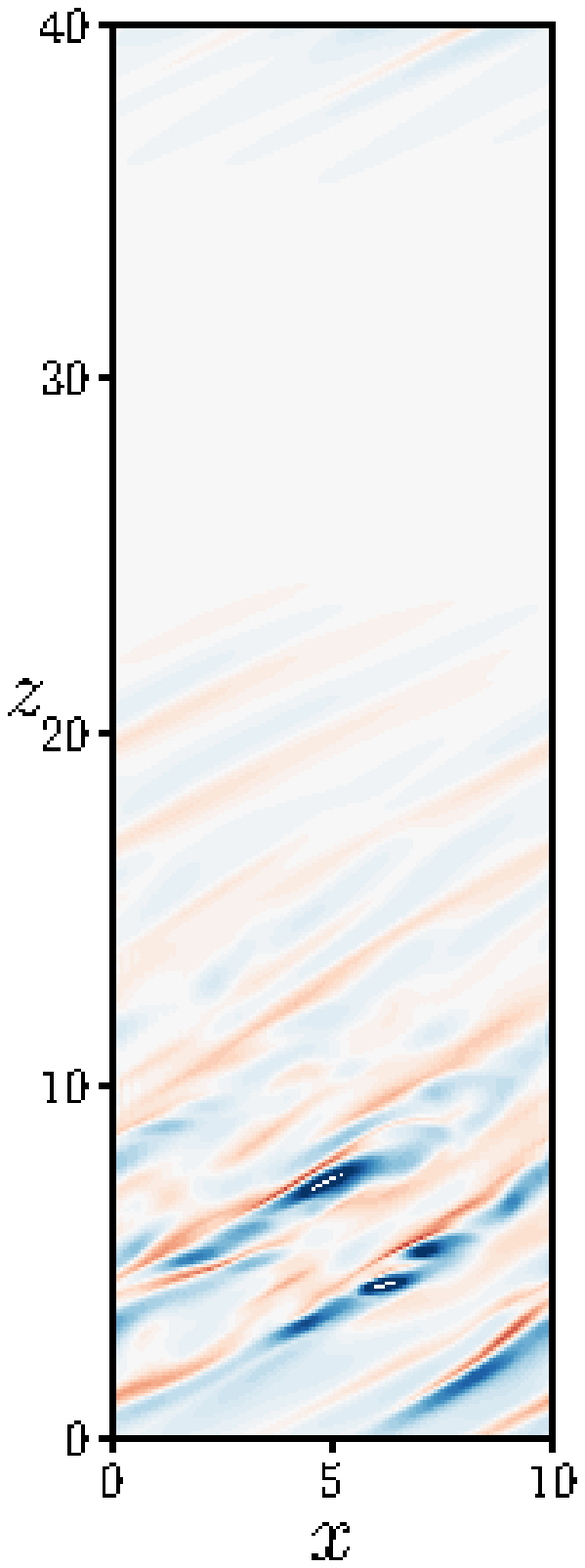}}
        \subfigure[$t=150$]{\includegraphics[width=.15\columnwidth]{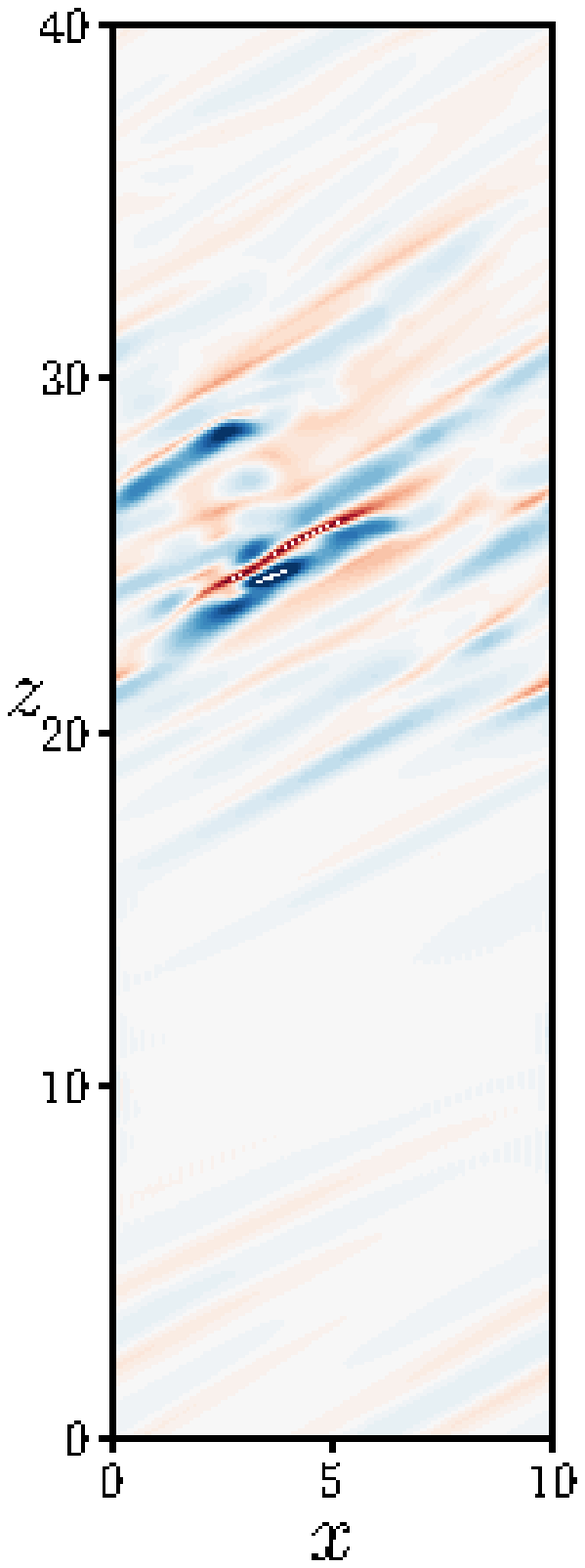}}
        \caption[]{Time evolution of the localised linear optimal perturbation at different times: shaded contours of the wall-normal velocity at $y=0.25$.} 
        %ENZA: che energia iniziale ha questa perturbazione?? Inoltre, dobbiamo aggiungere un'altra figura che fa vedere il calcolo nel dominio grande, fatto usando questa perturbazione ottimale localizzata, calcolata nel dominio tilted. 
        \label{locLOPevolution}
\end{figure}
This localised perturbation is injected in the DNS with different initial energies. Its time evolution for the minimal initial energy able to induce turbulent bands, i.e., $E_0=3.3 \times 10^{-3}$, is reported in figure \ref{locLOPevolution}. At first, the oblique streaks increase their amplitude ($t=20$) and start to saturate nonlinearly, until secondary instability arises ($t=50$) and triggers turbulence in a localised zone plunged in the laminar flow ($t=80-100$). At $t = 150$, the flow presents the same configuration shown in figure \ref{DNS} for a turbulent band generated by decreasing the Reynolds number starting from a fully turbulent velocity field. Notably, inflection points similar to those observed in figure \ref{1DtT_NLOP}, are observed at small time in the velocity profiles (see the dashed line on figure \ref{1DtT_LOP}). Thus, it appears that for triggering a turbulent band in the tilted domain, starting from a rather weak perturbation, two main elements are needed: small-scale oblique streaks aligned with the baseflow, that saturate creating inflection points, and a large-scale vortical flow ensuring spatial localisation in the $z$ direction. The transition at the small-scale is due to the classical lift-up mechanism, followed by secondary instability of the saturated streaks, which triggers the self-sustained cycle supporting turbulence \citep{Hamilton1995,Waleffe1997}. However, in the absence of a large-scale flow ensuring localisation and allowing to maintain the band, these mechanisms are not sufficient to generate localised turbulence. Of course the initial phase of growth due to the lift-up mechanism can be skipped by directly feeding the flow with inflection points, as done by \cite{song2020}, but at the cost of a larger amplitude disturbance, which can be more difficult and expensive to obtain in an experimental setup.\\
Finally, we should verify whether this artificially-localised perturbation able to optimally produce streaks, can generate turbulent bands also in large, non-tilted domains, where no angle is imposed a priori. Thus, we have reported the artificially-localised linear optimal perturbation computed in the tilted domain, in a very large (non-tilted) domain of size $L_{x'}\times L_{y'}\times L_{z'}=250\times2\times125$, and let it evolve freely by a DNS. 
As shown in figure \ref{locLOPevolution-large}, despite at $t=0$ a large-scale flow is present only in the vicinity of the perturbation, at $t=100$ a clear quadrupolar large-scale vortical structure, filling the whole domain, is observed. Notice that, as discussed in \cite{wang_duguet2020}, a quadrupolar structure arises in the presence of a  negative  spanwise vorticity generated near the walls inside a spot, as a consequence of the shearing of the streamwise velocity and the breaking of the spanwise homogeneity. New streaky structures, generated by the self-sustained process triggered by the optimal counter-rotating vortices and streaks, are then created following the shear layer between two of the previously observed large-scale vortices ($t=300$), finally creating a clear turbulent band ($t=800$).
Despite not being optimal for this large, non-tilted domain, this  perturbation is able to generate a large-scale flow that promotes the formation of small-scale streaks in an oblique direction,  consequently inducing the band formation. The optimization of perturbations in this large non-tilted domain is beyond the scope of the present work, and is treated in detail in \cite{ParenteJFM2021}, where the minimal-energy optimal perturbations able to generate turbulent bands are computed and discussed for different values of $Re$.
%\vspace{-0.5cm}
\section{Conclusion}\label{sec:conclusion}
In this work we have investigated the energy growth mechanisms involved in the laminar-turbulent transition in the form of turbulent bands using linear and nonlinear optimization. We have considered a plane Poiseuille flow at $Re = 1000$ in a tilted domain with angle $\theta = 35^{\circ}$ that exhibits a single turbulent band. Linear optimization have reported that the optimal perturbation is three-dimensional and aligned with the oblique baseflow, in the form of low- and high- speed streaks modulated in the streamwise and spanwise directions with $k_x =  1.2$ and $k_z=- 1.75$, respectively. Similar wavenumbers are found at the same Reynolds number by direct numerical simulation and by linear stability analysis at the head of the turbulent band, where an angle comparable to that of the optimal streaks is observed. %These values are in line with those reported by \cite{xiao2020} performing a l.
However, the linear optimal perturbation needs  a very large initial energy to trigger turbulence, which spreads in the whole domain. Using nonlinear optimization, a localized turbulent band is triggered for an initial energy five orders of magnitude weaker, $E_0 = 2.1 \times 10^{-5}$. The nonlinear optimal perturbation is characterised by a localised small-scale flow and a large-scale flow surrounding it. The small-scale flow is composed by oblique counter-rotating vortices and streaks with an angle comparable to that found via linear optimization, which develop inflection points at target time. 
 In order to isolate the influence of large-scale flow and localization from that of the small-scale structures, we have constructed a localized perturbation  by artificially confining the linear optimal to a localized region in the spanwise direction and injected it on the laminar flow both in the tilted and in a non-tilted, very large domain. In both domains, a turbulent band is created.
These results suggest that transition to a turbulent band might arise due to the optimal lift-up mechanism when coupled with a large-scale vortical flow intimately linked to the spatial localisation of the disturbance. This energy growth mechanism provides high-amplitude streaks developing inflection points when saturating nonlinearly, but since the optimal streaks are aligned with the base flow, they cannot generate a turbulent band by themselves. However, the large-scale flow generated by the spatial localisation of the perturbation provides the preferential direction of spreading of the streaks generated by the lift-up mechanism, and is thus necessary  to trigger turbulence in the form of turbulent bands. %Particularly, the small scale flow formation is linked to a linear mechanism, notably the classical lift-up mechanism that acts in the base flow direction. Instead, nonlinear effects are necessary to localise the perturbation and create a large scale flow.\\
%Furthermore, the oblique streaks might be linked to the oblique waves upstream the turbulent band, due to the similar characteristics in term of angle. Thus, these structures can be viewed as the low and speed streaks observed in both tilted and large domains responsible of the turbulent band growth.

%\backsection[Supplementary data]{\label{SupMat}Supplementary material and movies are available at \\https://doi.org/10.1017/jfm.2019...}

%\backsection[Acknowledgements]{Acknowledgements may be included at the end of the paper, before the References section or any appendices. Several anonymous individuals are thanked for contributions to these instructions.}

\backsection[Funding]{This  work  was  granted  access  to  the  HPC  resources  of  IDRIS  under  the  allocation 2020- A0072A06362 and A0092A06362 made by GENCI.}

\backsection[Declaration of interests] 
{The authors report no conflict of interest.}
%\vspace{-0.5cm}
%\backsection[Data availability statement]{The data that support the findings of this study are openly available in [repository name] at http://doi.org/[doi], reference number [reference number].}

%\backsection[Author ORCID]{Authors may include the ORCID identifers as follows.  F. Smith, https://orcid.org/0000-0001-2345-6789; B. Jones, https://orcid.org/0000-0009-8765-4321}

%\backsection[Author contributions]{Authors may include details of the contributions made by each author to the manuscript, for example, ``A.G. and T.F. derived the theory and T.F. and T.D. performed the simulations.  All authors contributed equally to analysing data and reaching conclusions, and in writing the paper.''}

\bibliographystyle{jfm}
%\bibliography{jfm}
\bibliography{biblio.bib}

\end{document}